\documentclass[prc,preprint,showpacs,showkeys,lengthcheck,nofootinbib,tightenlines,
  twocolumn,notitlepage,preprintnumbers,superscriptaddress]{revtex4-1}

\usepackage[utf8]{inputenc}
\usepackage{newtxtext}
\usepackage[libertine]{newtxmath}

\usepackage{float}
\usepackage{graphicx}
\usepackage[caption=false]{subfig}
\usepackage[usenames,dvipsnames]{xcolor}
\graphicspath{{figures/}}

\usepackage{hyperref}
\hypersetup{colorlinks=true,citecolor=blue,linkcolor=blue,urlcolor=blue}

\usepackage{diagbox}
\usepackage{booktabs}

\usepackage{slashed}
\usepackage{bm}

\usepackage{outlines}
\usepackage{enumitem}
\setenumerate[1]{label=\Roman*.}
\setenumerate[2]{label=\Alph*.}
\setenumerate[3]{label=\roman*.}
\setenumerate[4]{label=\alph*.}

\usepackage{titlesec}
\titlespacing{\section}{4pt}{10pt plus 4pt minus 2pt}{8pt plus 2pt minus 2pt}
\def\hs{\hspace}

\def\no{\nonumber}

\def\lf{\left}
\def\rg{\right}

\newcommand{\emdash}{\hs{1pt}---\hs{1pt}}

\begin{document}

\title{Intrinsic Glue and Wilson lines within Dressed Quarks}

\author{Caroline~S.~R. Costa}
\email{csr.costa@unesp.br}
\address{Instituto de F\'{i}sica Te\'{o}rica, Universidade Estadual Paulista, Rua Dr.  Bento Teobaldo Ferraz, 271 - Bloco II, 01140-070,
S\~{a}o Paulo, SP, Brazil}

\author{Adam Freese}
\email{afreese@uw.edu}
\address{Physics Division, Argonne National Laboratory, Lemont, Illinois 60439, USA}
\address{Department of Physics, University of Washington, Seattle, WA 98195, USA}

\author{Ian C. Clo\"{e}t}
%\email{icloet@anl.gov}
\address{Physics Division, Argonne National Laboratory, Lemont, Illinois 60439, USA}

\author{Bruno El-Bennich}
\address{Laborat\'orio de F\'isica Te\'orica e Computacional, Universidade Cidade de S\~ao Paulo,
Rua Galv\~ao Bueno 868, 01506-000, S\~ao Paulo, SP, Brazil}

\author{Gast\~ao Krein}
\address{Instituto de F\'{i}sica Te\'{o}rica, Universidade Estadual Paulista, Rua Dr.  Bento Teobaldo Ferraz, 271 - Bloco II, 01140-070,
S\~{a}o Paulo, SP, Brazil}

\author{Peter C. Tandy}
\address{Center for Nuclear Research, Department of Physics, Kent State University, Kent OH 44242, USA}
\address{CSSM, Department of Physics, University of Adelaide, Adelaide SA 5005, Australia}

\begin{abstract}
We construct a quark target model (QTM) to incorporate intrinsic glue into effective low-energy models of QCD, which often contain only quark degrees of freedom. This method guarantees the gauge invariance of observables order-by-order in the strong coupling. 
%The QTM is solved at leading order, and the quark and gluon PDFs for the dressed quarks are obtained.
The quark and gluon PDFs for the dressed quarks are obtained in the QTM at leading order.
We demonstrate gauge invariance of the results by comparing both covariant and light cone gauges, with the former including an explicit Wilson line contribution. A key finding is that in covariant gauges the Wilson line can carry a significant amount of the light cone momentum. With coupling strength $\alpha_s = 0.5$ and dressed quark mass $M_q = 0.4\,$GeV, we find quark and gluon momentum fractions of $\lf<x\rg>_q = 0.81$ and  $\lf<x\rg>_g = 0.19$, where the Wilson line contribution to the quark momentum fraction is $-0.18$. We use the on-shell renormalization scheme and find that at one-loop this Wilson line contribution does not depend on the covariant gauge but does vanish in light cone gauge as expected. This result demonstrates that it is crucial to account for Wilson line contributions when calculating quantum correlation functions in covariant gauges. We also consider the impact of a gluon mass using the gauge invariant formalism proposed by Cornwall, and combine these QTM results with two quark-level models to obtain quark and gluon PDFs for the pion.
%These QTM results can be combined with many effective models for hadron structure. In this work we obtain results for the quark and gluon PDFs of the pion, by combining the QTM results with both a phenomenological model for the pion based on QCD asymptotics, and the model of Nambu and Jona-Lasinio. These results are 
\end{abstract}

\maketitle

%%%%%%%%%%%%%%%%%%%%%%%%%%%%%%%%%%%%%%%%%%%%%%%%%%%%%%%%%%%%%%%%%%%%%%%%%%%%%%%%
\section{INTRODUCTION}
Many of the open mysteries in nuclear physics\emdash such as the origin of hadron mass and the distribution of spin in the proton\emdash can be addressed through partonic correlation functions, including parton distribution functions (PDFs), transverse momentum distributions (TMDs), and generalized parton distributions (GPDs). However, calculating these distributions exactly from quantum chromodynamics (QCD) remains challenging, where even in lattice QCD approximation schemes are needed~\cite{Ji:2013dva,Radyushkin:2017cyf,Ma:2017pxb,Izubuchi:2018srq,Sufian:2020vzb,Joo:2020spy,Gao:2020ito}. In this milieu model calculations, such as, the Dyson-Schwinger equations (DSE)~\cite{Cloet:2013jya,Eichmann:2016yit}, the chiral quark soliton model~\cite{Diakonov:1996sr,Diakonov:1997vc,Diakonov:1998ze}, and the Nambu--Jona-Lasinio (NJL) model~\cite{Cloet:2014rja,Freese:2019eww} can offer important insight.

The primary emphasis of model calculations has so far been on calculating various quark correlation functions, with information about the gluons either inferred indirectly or generated entirely perturbatively by QCD evolution equations~\cite{Bednar:2018mtf,Shi:2020pqe,Cui:2020tdf,Freese:2021zne}. Such approaches are not adequate to fully address several of the major open questions in hadron physics, which involve significant intrinsic gluonic contributions at all renormalization scales. In addition, a major focus of future experimental efforts such as at the Electron-Ion Collider is the gluon structure of hadrons and nuclei. It is therefore vital that gluonic observables be directly calculated in effective models of QCD.

The goal of this work is to consider one avenue for direct calculation of gluon observables. Specifically, we take the approach of adding intrinsic gluons to effective models of QCD that involve only quark degrees of freedom, such as the NJL model. Since these models have been successful in the calculation of observables such as PDFs~\cite{Cloet:2007em,Cloet:2005pp}, form factors~\cite{Cloet:2014rja,Hutauruk:2018zfk}, and GPDs~\cite{Freese:2019eww,Freese:2020mcx}, this success can be carried over without constructing a new effective model whole cloth.

Perhaps the most straightforward method of adding intrinsic glue to an effective model with quark degrees of freedom is to simply resolve the gluonic substructure of the dressed quarks. These dressed quarks are a shared feature of many low-energy models of QCD, and are emergent effective degrees of freedom (massive quasi-particles) with the same quantum numbers as the almost massless current quarks appearing in the QCD Lagrangian. These dressed quarks are made up of many quarks, anti-quarks, and gluons, and are connected to dynamical chiral symmetry breaking in QCD. 

In this work, we construct a quark target model (QTM) to calculate the quark and gluon substructure of a dressed quark to leading order in the quark-gluon coupling strength $(\alpha_s)$. We focus on the quark and gluon PDFs of the quark target. In general gauges, it is necessary to account for Wilson line contributions to the quark PDF in order to respect gauge invariance and to satisfy the momentum sum rule.\footnote{Beyond leading order there is also a Wilson line contribution to the gluon PDF but it does not contribute to the gluon light cone momentum.} For example, within the Dyson-Schwinger equation approach PDFs have been calculated using Landau gauge~\cite{Cui:2020tdf,Freese:2021zne}, however, these results ignore the Wilson line contributions that must be present in covariant gauges. We explore covariant and light cone gauges to explicitly demonstrate the gauge independence of the QTM quark and gluon PDFs, and the importance of the Wilson line contribution in covariant gauges.

This work is organized as follows: In Sec.~\ref{sec:qtm}, we construct the QTM and obtain results for the quark and gluon PDFs of the quark target. Numerical results for the PDFs and their Mellin moments are presented for massive quarks and massless gluons. In Sec.~\ref{sec:gmass}, an explicit gluon mass is introduced and its impact on the QTM PDFs is studied under two scenarios, one where a naive mass term is added to the Lagrangian which directly breaks gauge invariance, and in the other scenario we use the formalism proposed by Cornwall~\cite{Cornwall:1981zr} that maintains gauge invariance and the momentum sum rule via the introduction of an auxiliary field. In Sec.~\ref{sec:pion}, we combine the QTM PDFs with two quark-only pion PDFs models, thereby obtaining quark and gluon PDFs for the pion at the model scale. DGLAP evolution is then performed and the PDFs are compared to data and an empirical PDF parameterization. Finally, in Sec.~\ref{sec:conc}, we provide some conclusions and an outlook.

%%%%%%%%%%%%%%%%%%%%%%%%%%%%%%%%%%%%%%%%%%%%%%%%%%%%%%%%%%%%%%%%%%%%%%%%%%%%%%%%
\section{QUARK TARGET MODEL\label{sec:qtm}}
We consider a QTM, which consists of taking the target state to be a dressed quark, and use this model to compute its quark and gluon PDFs at one-loop in perturbative QCD. Although quarks are not asymptotic states in QCD, a QTM describes the properties of quarks through matrix elements of operators between on-shell quark states, denoted by $\lf|Q(P)\rg>$. In the following we formulate the QTM and present the definitions of the PDFs of interest.

%%%%%%%%%%%%%%%%%%%%%%%%%%%%%%%%%%%%%%%%%%%%%%%%%%%%%%%%%%%%%%%%%%%%%%%%%%%%%%%%
\subsection{Definitions and Feynman Rules}
To formulate the QTM we begin with an effective quark-gluon Lagrangian of the form:
\begin{equation}
    \label{QCD_lagrangian}
    \mathcal{L}
    =
    \sum_{q}\bar{\psi}_{q}(i\slashed{D} - m_{q})\psi_{q}
    - \frac{1}{4}G_{\mu\nu}^{a}G^{\mu\nu}_{a}
    +
    \mathcal{L}_{\mathrm{GF}},
\end{equation}
where $\psi_q$ denotes a quark field with mass $m_q$, and a gauge-fixing term $\mathcal{L}_{\mathrm{GF}}$ is necessary to quantize the theory and fully define the gluon propagator~\cite{Collins:2011zzd}. We emphasize that the quark and gluon fields appearing in this QTM are not the current quark and gluon fields appearing in the QCD Lagrangian. The QTM fields are dressed by the interactions of an underlying low-energy effective theory of QCD, and therefore $m_q$ is not the current quark mass. In principle, the gluons appearing within the dressed quarks can also be dressed and behave as if they have a dynamically generated infra-red scale~\cite{Falcao:2020vyr}. Nevertheless, within the context of the QTM, we shall refer to $\psi_q$ and $m_q$ as the ``bare quark field'' and ``bare quark mass'', respectively. We limit calculations within the QTM to $O(g^2)$, as such, there are no gluon self-interactions in this work and the leading-order QTM is effectively abelian. Results presented in this section are therefore leading order in $\alpha_s = g^2/(4\pi)$. 

A quark target of flavor $Q$ (capital letters are used to denote dressed quarks to distinguish them from bare quarks) has an unpolarized quark distribution defined by
\begin{align}
f_{q/Q}(x) &= \int\frac{\mathrm{d}\,\lambda}{4\,\pi}\, e^{ix\lambda\,P\cdot n} \no \\
&\hs*{7mm}
\times
\lf< {\rm Q}(P)\rg|\bar{\psi}_{q}(0)\,\slashed{n}\, W(0,n\lambda)\,\psi_{q}(n\lambda)\lf|{\rm Q}(P)\rg>,
\label{def-fqQ}
\end{align}
where $n$ is a light-like vector defining the light cone and $x=k\cdot n/P\cdot n$ is the light cone momentum fraction carried by the struck bare quark. The Wilson line operator, $W(0, n\lambda)$, which connects the bare quark fields and ensures the color gauge invariance of the PDF reads
\begin{equation}
  \label{wilson-line}
  W(0, n\lambda) = {\rm P} \, e^{-ig\int_{\lambda}^{0}\mathrm{d}\xi\, n\cdot A(n\xi)},
\end{equation}
with ${\rm P}$ denoting the path-ordering operator. The unpolarized gluon distribution in the quark target is defined by
\begin{align}
    xf_{g/Q}(x)
    &= \frac{1}{P\cdot n}
    \int\frac{\mathrm{d}\lambda}{2\pi}\,
    e^{ix\lambda\,P\cdot n} \no \\
&\hs*{4mm}
\times
\langle Q(P)| G^{+\,}_{\,\mu}(0)\,W_A(0, n\lambda)\,G^{\mu +}(n\lambda)|Q(P)\rangle,
    \label{def-fgQ}
\end{align}
where $W_A(0, n\lambda)$ is a Wilson line in the adjoint representation.

The quark and gluon PDFs in the QTM are calculated by evaluating Feynman diagrams. The momentum space Feynman rules for the operators defining the PDFs are given by:
\begin{align}
\label{pdf_qq}
\vcenter{\hbox{\includegraphics[scale=0.6]{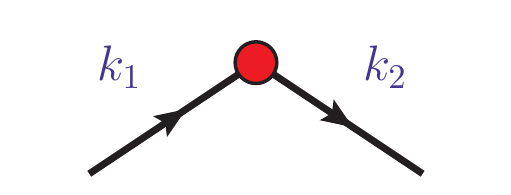}}}
  &=
  \frac{\slashed{n}}{2}\ 
  \delta\left(n\left[xP - \frac{k_1+k_2}{2}\right]\right), \\
\label{pdf_gg}
\vcenter{ \hbox{ \includegraphics[scale=0.6]{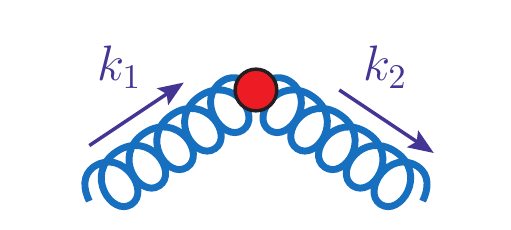} } }
  &=
  \frac{-1}{P\cdot n}\,\delta_{a_1 a_2}
  \left[ (nk_{1})g^{\rho\mu_1}-k_1^{\rho}n^{\mu_1} \right]
  \nonumber \\[-1.2em]
& \hspace*{15.5mm}
\times \left[ (nk_{2})g_{\rho \,}^{\,\mu_2}-k_{2\rho}n^{\mu_2}\right]
\nonumber \\
& \hspace*{-26mm} \times
\left\{
\delta\left(n\left[ xP - \frac{k_1 + k_2}{2}\right] \right)
+ \delta \left(n\left[ xP +\frac{k_1 + k_2}{2}\right] \right)\right\}, \\
\label{pdf_qqg}
\vcenter{  \hbox{ \includegraphics[scale=0.6]{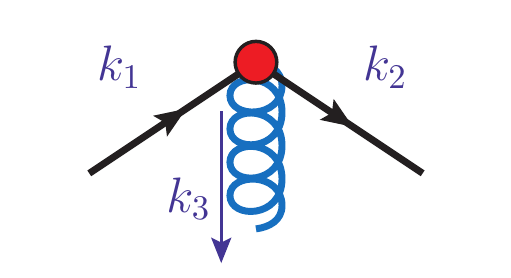} } }
   &=
  -\frac{g\,n^{\mu}}{k_3\cdot n}\frac{\slashed{n}\,t^a}{2}
    \nonumber
    \\
&\hs*{-32mm} 
\left\{
    \delta \left(n\left[ xP - \frac{k_1 + k_2+k_3}{2}\right] \right)
 - \delta \left(n\left[ xP - \frac{k_1 + k_2-k_3}{2}\right] \right)
    \right\},
\end{align}
where the operator insertion (large dot) can be on a quark line, a gluon line, or a quark-quark-gluon vertex which results from the Wilson line. The Wilson line Feynman rule  can be derived by first splitting the Wilson line appearing in Eq.~(\ref{def-fqQ}) into two pieces, one going to infinity and the other back to the endpoint:
\begin{align}
    W(x,y)
    &= W(x,+\infty) W(+\infty,y)
    \no
    \\
    &=[W(+\infty, x)]^{\dagger}W(+\infty,y).
\label{wilson_line_2}
\end{align}
Expanding each of the Wilson lines to the first order in $g$ gives
\begin{align}
    [W(+\infty, x)]^{\dagger}  W(+\infty,y)
     &=
    1
    + ig\int_{x}^{\infty} \mathrm{d}\xi\ n\cdot A(\xi)
    \no
    \\
    &- ig\int_{y}^{\infty}\mathrm{d}\xi\ n\cdot A(\xi)
    + O(g^2).
    \label{wilson_line_3}
\end{align}
When this expression is inserted in Eq.~\eqref{def-fqQ}, the first term gives the diagram in Eq.~(\ref{pdf_qq}), while the two order-$g$ terms are responsible for the Wilson line diagram in Eq.~(\ref{pdf_qqg}). The gluon Feynman rule can be derived by writing the gluon field strength tensor appearing in Eq.~(\ref{def-fgQ}) in terms of the gluon field and expanding to zeroth order in $g$.

%===============================================================================
%===============================================================================
\subsection{Quark and Gluon Distributions in QTM}
\begin{figure}[tbp]
\centering\includegraphics[width=\columnwidth]{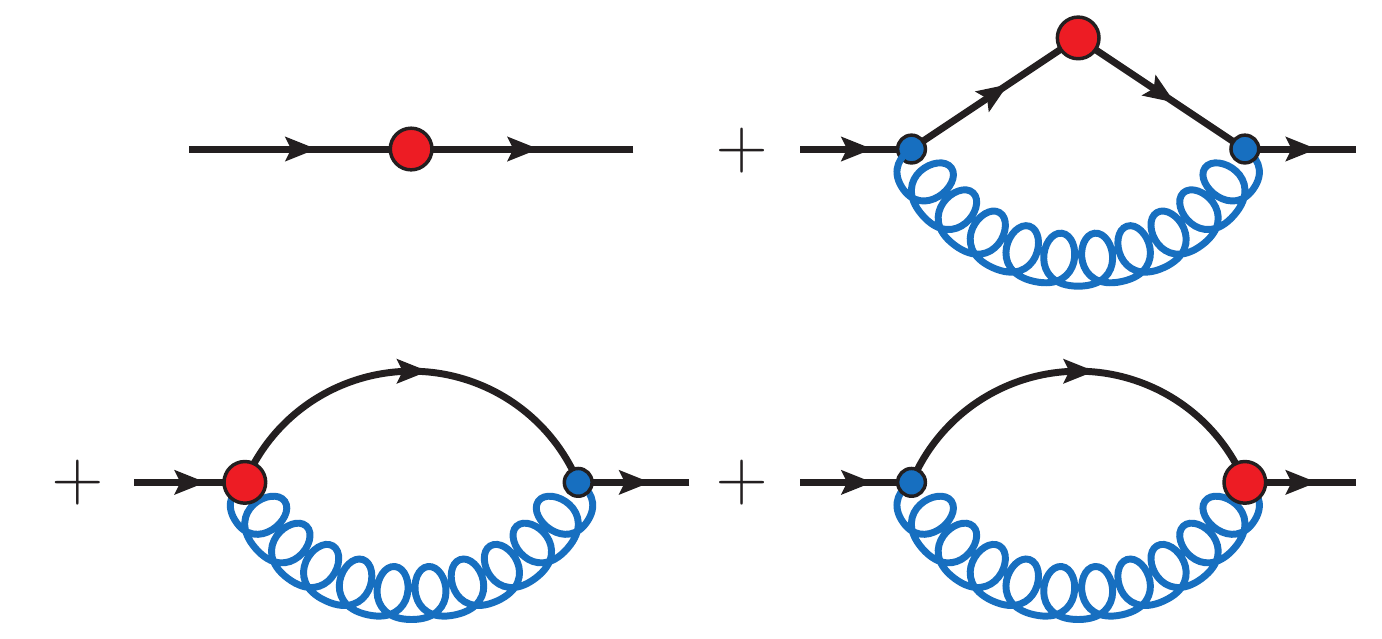}
\caption{Leading order diagrams contributing to the quark PDF in the quark target model. The solid lines represent a quark propagator and the curly lines a gluon propagator. The large vertices are given by the Feynman rules of Eqs.~\eqref{pdf_qq}--\eqref{pdf_qqg}, and the small vertex is the standard quark-gluon interaction obtained from the QCD Lagrangian.}
\label{fig:qtm_pdf_q}
\end{figure}

\begin{figure}[tbp]
\centering\includegraphics[scale=0.75]{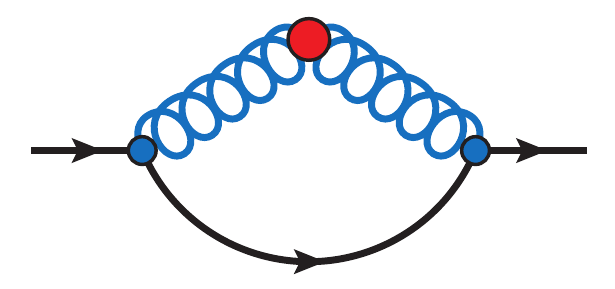}
\caption{Leading order diagram contributing to the gluon PDF in the quark target model. The solid line represents a quark propagator and the curly lines a gluon propagator. The large vertex are given by the Feynman rule in Eq.~\eqref{pdf_gg}, and the small vertices is the standard quark-gluon interaction obtained from the QCD Lagrangian.}
\label{fig:qtm_g_gluon}
\end{figure}

The one-loop diagrams contributing to the quark PDF of the quark target are depicted in Fig.~\ref{fig:qtm_pdf_q} and the quark target gluon PDF is given by the single diagram in Fig.~\ref{fig:qtm_g_gluon}. Adding all four diagrams in Fig.~\ref{fig:qtm_pdf_q}, the quark PDF can be written as
\begin{equation}
f_{q/Q}(x) = Z_2\,\delta(1-x) + f^{\mathrm{tri}}_{q/Q}(x) + f^{W}_{q/Q}(x),
\label{eqn:fqQ}
\end{equation}
where the two diagrams that represent the Wilson line are equal, and have been included as a single contribution in Eq.~\eqref{eqn:fqQ}. The quark field renormalization constant, $Z_2$, is implicitly present in every diagram, since the operator defining the PDFs is given in terms of the unrenormalized fields~\cite{Collins:2011zzd}. However, $Z_2$ can also be expanded as a series in $g$, with the leading-order form $Z_2 = 1 + g^2 Z_2^{(2)} + O(g^3)$, thus, in a leading-order calculation $Z_2$ effectively only contributes to the first diagram in Fig.~\ref{fig:qtm_pdf_q}. Note, the PDF definitions given in Eqs.~\eqref{def-fqQ} and \eqref{def-fgQ} are gauge invariant, but the individual diagrams in Fig.~\ref{fig:qtm_pdf_q} need not be. For example, in light cone gauge $n\cdot A=0$ and therefore the Wilson line contributions vanish, which is not true in general.

%===========================================================
%===========================================================

The second diagram in Fig.~\ref{fig:qtm_pdf_q} (triangle diagram) reads
\begin{align}
  f^{\mathrm{tri}}_{q/Q}(x)
  &= \frac{ig^2}{2}\frac{C_{F}}{P\cdot n}
  \int \frac{\mathrm{d}^{4}k}{(2\pi)^{4}}\
  \delta\left(x-\frac{k\cdot n}{P\cdot n}\right)
  \no
  \\
  &\hs*{5mm}
\times\, \bar{u}(P)\,\gamma^{\mu}\,S(k)\,\slashed{n}\,S(k)\,\gamma^{\nu}\,D_{\mu\nu}(P-k)\,u(P),
  \label{qq_pdf}
\end{align}
where $S(k)$ is the quark propagator and $D_{\mu\nu}(k)$ the gluon propagator. The order-$g$ Wilson line terms give rise to the third and fourth diagrams in Fig.~\ref{fig:qtm_pdf_q}. These diagrams are equal and together are given by
\begin{align}
f^{W}_{q/Q}(x)
     &=
     g^2\,\frac{C_F}{P\cdot n}\,n^{\nu}\!
     \int \frac{\mathrm{d}^{4}k}{(2\pi)^{4}}\!
     \left[
       \delta\left(x-\frac{k\cdot n}{P\cdot n}\right)
       -\delta(1-x)\right] \no \\
&\hs*{-6mm}\times\, \frac{i}{n\cdot(P-k)+i0}\ 
     \bar{u}(P)\,\gamma^{\mu}\,S(k)\,\slashed{n}\,D_{\mu\nu}(P-k)\,u(p).
\end{align}
Note, the term $1/(n\cdot k + i0)$ is part of the operator in this Feynman diagram approach, however it is connected to the eikonal propagator piece of the Wilson line in the cut diagram method~\cite{Collins:2011zzd}. The gluon PDF of the quark target, given by the diagram in Fig.~{\ref{fig:qtm_g_gluon}}, reads
\begin{align}
   & x\,f_{g/Q}(x) 
    =
    -i\, g^2\, \frac{C_F}{P\cdot n}
    \int\frac{\mathrm{d}^{4}k}{(2\pi)^{4}}\
    \bar{u}(P)\, \gamma ^{\mu}\,S(P-k)\,\gamma ^{\nu}\,u(P)
    \no
    \\
    &\hs*{4mm} \times
    \big[n\cdot k\,g_{\rho}^{\alpha}-k_{\rho}n^{\alpha}\big]
    \big[n\cdot k\,g^{\rho\beta}-k^{\rho}n^{\beta}\big]
       \no
      \\
      &\hs*{4mm} \times \lf[ \delta\left(n\left[xP-k\right] \right)
      + \delta\left(n\left[xP+k\right] \right)
      \rg] D_{\mu\alpha}(k)\,D_{\nu\beta}(k).
\end{align}
These expressions for the PDFs are evaluated by first taking Mellin moments to eliminate the delta function, and then writing them in a form where the inverse Mellin transform is trivial. 

At order $g^2$ in the QTM the gluon propagator does not get dressed, therefore, in a general covariant gauge and in light cone gauge the gluon propagators take the standard form:
\begin{subequations}
  \begin{align}
    D_{\mu\nu}^{\mathrm{cov}}(k) 
    &=
    \frac{-1}{k^2+i0}\lf[g_{\mu\nu} - (1-\xi)\frac{k_\mu k_\nu}{k^2+i0}\rg],
    \\
    D^{\mathrm{LC}}_{\mu\nu}(k) 
    &=
    \frac{-1}{k^2+i0} \left[
      g_{\mu\nu} - \frac{k_\mu n_\nu + n_\mu k_\nu}{(k\cdot n)}
      \right].
  \end{align}
\end{subequations}
At order $g^2$ the quark propagator does get dressed, where the self-energy reads
\begin{equation}
\Sigma(p) = ig^2\,C_{F}\int \frac{\mathrm{d}^4 k}{(2\pi)^4}\ \gamma^\mu\,S(p-k)\,\gamma^\nu\,D_{\mu\nu}(k),
\end{equation}
and takes the general form
\begin{equation}
    \Sigma(p) = A(p^2)\slashed{p}+B(p^2)+ C(p^2)\slashed{n},
\end{equation}
where in covariant gauges $C(p^2)=0$. The quark field renormalization constant, $Z_2$, is given by
\begin{equation}
 Z_{2}^{-1}
  =
  \left.
  1-\frac{\mathrm{d}\,\Sigma(p)}{\mathrm{d}\slashed{p}}
  \right|_{\slashed{p}=M_{q}}.   
\label{eq:Z2}
\end{equation}
where $M_q$ is the renormalized (physical) dressed quark mass.

%===========================================================
%===========================================================

The physical dressed quark mass and quark field renormalization constant can be understood by considering the gap equation for the \emph{unrenormalized} quark propagator:\footnote{
We emphasize that dressing and renormalization are distinct procedures. Dressing accounts for the incorporation of interactions in the calculation of Green's functions, whereas renormalization rescales the fields through factors such as $Z_2$. These procedures are often done in tandem since\emdash when an infinite UV regulator is taken\emdash the dressing produces UV divergences that must be contained by renormalization. The unrenormalized and renormalized propagator are both dressed: they are defined respectively as time-ordered two-point Green's functions of the unrenormalized and renormalized quark fields and differ by a factor $Z_2$, i.e., $S_0(p) = Z_2\,S(p)$.
}
\begin{equation}
    S_0^{-1}(p) = \slashed{p} - m_q - \Sigma(p),
\end{equation}
where $S_0(p)$ is the unrenormalized dressed quark propagator,
$m_q$ is the bare quark mass, and $\Sigma(p)$ is the self-energy.
The self-energy, which accounts for changes in the quark propagator
due to the quark and gluons interactions,
shifts the pole of the propagator to
what we identify as being the physical mass $M_{q}$:
\begin{equation}
    M_{q} = m_{q}+\Sigma(\slashed{p}=M_{q}).
\end{equation}
The shift of the quark mass also causes the residue of the propagator to change.
Near the physical pole $\slashed{p} \simeq M_{q}$,
\begin{equation}
S_0(p) =
    \frac{1}{\slashed{p}-m_{q}-\Sigma(p)}
    \simeq \frac{Z_2}{\slashed{p}-M_{q}},
\end{equation}
where $Z_2$ is the field renormalization constant of Eq.~\eqref{eq:Z2}.

To study the gauge dependence of the various contributions to the quark PDF of the quark target, given by the diagrams in Fig.~\ref{fig:qtm_pdf_q}, we explicitly consider general covariant and light cone gauges. For the quark field renormalization constant we find
\begin{align}
\label{eq:Z2cov}
Z_2^{\mathrm{cov}} &= 1 + 2\,g^2\,C_F
    \int_{0}^{1}\mathrm{d}x \no \\ 
&\hs*{0mm}
\left[x\,I_{2}(w_q(x)) + 4\,M_{q}^2\,x(1-x)(2-x)\,I_3(w_q(x))\right], \\
\label{eq:Z2LC}
Z_{2}^{\mathrm{LC}} &= Z_{2}^{\mathrm{cov}} +
4\,g^2 C_F \int_0^1 \mathrm{d}x \ \frac{x}{1-x}\  I_2\big(w_q(x)\big),
\end{align}
where $w_q(x) = (1-x)^2 M_q^2$ is the Feynman mass parameter. We are using the on-shell renormalization scheme and therefore $Z_2^{\mathrm{cov}}$ is independent of the choice of covariant gauge~\cite{Grozin:2005yg}. The additional term in the LC result comes from the $n$ dependent piece of the light cone gauge gluon propagator. We employ a regularization scheme independent notation for the results where details are given in App.~\ref{sec:integrals}. 

The covariant and light cone gauge results for the triangle diagram contribution to the quark PDF of the quark target are
\begin{align}
\label{eq:tri_Feyn}
f^{\mathrm{tri,\,cov}}_{q/Q}(x) &=
  -2\,g^2C_{F}\Big[
     (1-x)\, I_{2}(w_{q}(x)) \no \\
&\hs*{22mm}
    + 4\,M_q^2\,x(1-x)\,
  I_{3}(w_{q}(x))
    \Big], \\
f^{\mathrm{tri},\mathrm{LC}}_{q/Q}(x) &=
  f^{\mathrm{tri,\,cov}}_{q/Q}(x)
  -
  4\,g^2C_{F}\, \frac{x}{1-x}\, I_{2}(w_{q}(x)),
  \label{eq:tri_LC}
\end{align}
where again we find that the triangle diagram in covariant gauges is independent of the gauge parameter $\xi$. Finally, the Wilson line contribution to the quark PDF is given by
\begin{align}
    \label{eqn:fWqQ}
f^{W,\mathrm{cov}}_{q/Q}(x) &= 4\,g^2\,C_{F}
\bigg[ \frac{x}{1-x}\,I_{2}(w_{q}(x)) \no \\
&\hs*{8mm} 
- \delta(1-x) \int_{0}^{1}\mathrm{d}y\, \frac{y}{1-y}\,I_{2}(w_{q}(y)) \bigg], \\
f^{W,\mathrm{LC}}_{q/Q}(x) &= 0,
\end{align}
where the light cone gauge result vanishes as expected and the covariant gauge result is independent of $\xi$. The appearance of a delta function term in the covariant gauge result corresponds to the so-called virtual diagram that appears in the cut diagram approach~\cite{Collins:2011zzd}. It has the same divergent behavior necessary to cancel the divergence in the first term. The gluon PDF for the quark target, at order $g^2$, does not have a Wilson line contribution and reads
\begin{align}
x f_{g/Q}(x) &= -2\,g^2C_{F}\Big[
\left[1+(1-x)^2 \right] I_{2}(w_g(x)) \no \\
&\hs*{23mm}
+ 4\,M_q^2\,x^2(1-x)\, I_3(w_g(x))\Big],
\label{eqn:fgQ}
\end{align}
where $w_g(x) = x^2 M_q^2 = w_q(1-x)$. Explicit calculation in any covariant gauge or light cone gauge gives the same result, as must be the case because $f_{g/Q}(x)$ is a gauge invariant quantity. 

Comparing the covariant and light cone gauge results for the quark PDF we find
\begin{align}
&Z_{2}^{\mathrm{LC}}\, \delta(1-x) + f^{\mathrm{tri},\mathrm{LC}}_{q/Q}(x) \no \\
&\hs*{19mm}
= Z_{2}^{\mathrm{cov}}\, \delta(1-x)
  +
  f^{\mathrm{tri},\mathrm{cov}}_{q/Q}(x)
  +
  f^{W,\mathrm{cov}}_{q/Q}(x),
\end{align}
as expected by gauge invariance of the PDF. So while the Wilson line is trivial in light cone gauge, covariant gauge  Wilson line contributions must appear elsewhere and in this calculation are split between $Z_2^{\mathrm{LC}}$ and $f^{\mathrm{tri},\mathrm{LC}}_{q/Q}(x)$.

%===============================================================================
%===============================================================================
\subsection{Wilson Line Contribution to PDF Sum Rules}
The quark number and momentum sum rules for the QTM can respectively be stated as:
\begin{align}
  \int_{0}^{1}\mathrm{d}x\, f_{q/Q}(x) &= 1, \\
\int_0^1\mathrm{d}x\left[x\,f_{q/Q}(x) + x\,f_{g/Q}(x)\right] &= 1.
\end{align}
It can be shown analytically that the quark PDF in Eq.~\eqref{eqn:fqQ} satisfies the quark number sum rule:
\begin{align}
 \int_{0}^{1}\mathrm{d}x\, f_{q/Q}(x) &=Z_2^{\mathrm{cov}}
  +
  \int_0^1\mathrm{d}x\, \lf[f_{q/Q}^{\mathrm{tri,\,cov}}(x) + f_{q/Q}^{\mathrm{W,\,cov}}(x)\rg]
  \no
  \\
&\hs*{-17mm}
=
  1
  +
  2\,g^2\, C_F \int_0^1 \mathrm{d}x\,
  \frac{\mathrm{d}}{\mathrm{d}x}
  \left[
    x(x-1) I_2\big(w_q(x)\big)
    \right]
  =
  1,
\end{align}
where the Wilson line does not contribute to the quark number sum rule as expected. However, this is not the case for the momentum sum rule, where in general the Wilson line does carry quark momentum. (We will show this explicitly in the next subsection.) Indeed, only when the Wilson line contribution is included do we find
\begin{align}
  f_{q/Q}(1-x) = f_{g/Q}(x) + \left[\delta(x)~\mathrm{terms}\right].
\end{align}
This symmetry ensures the momentum sum rule when the quark number sum rule is already satisfied, as it entails:
\begin{align}
  \int_{0}^{1}\mathrm{d}x \, x \left[f_{q/Q}(x)+f_{g/Q}(x)\right]
  =
  \int_{0}^{1}\mathrm{d}x \, f_{q/Q}(x)
  =
  1.
\end{align}
The fact that the Wilson line contributes to the momentum sum rule has important implications for the calculation of partonic correlation functions in QCD effective theories.

%===============================================================================
\begin{figure}[tbp]
  \centering
  \includegraphics[width=\columnwidth]{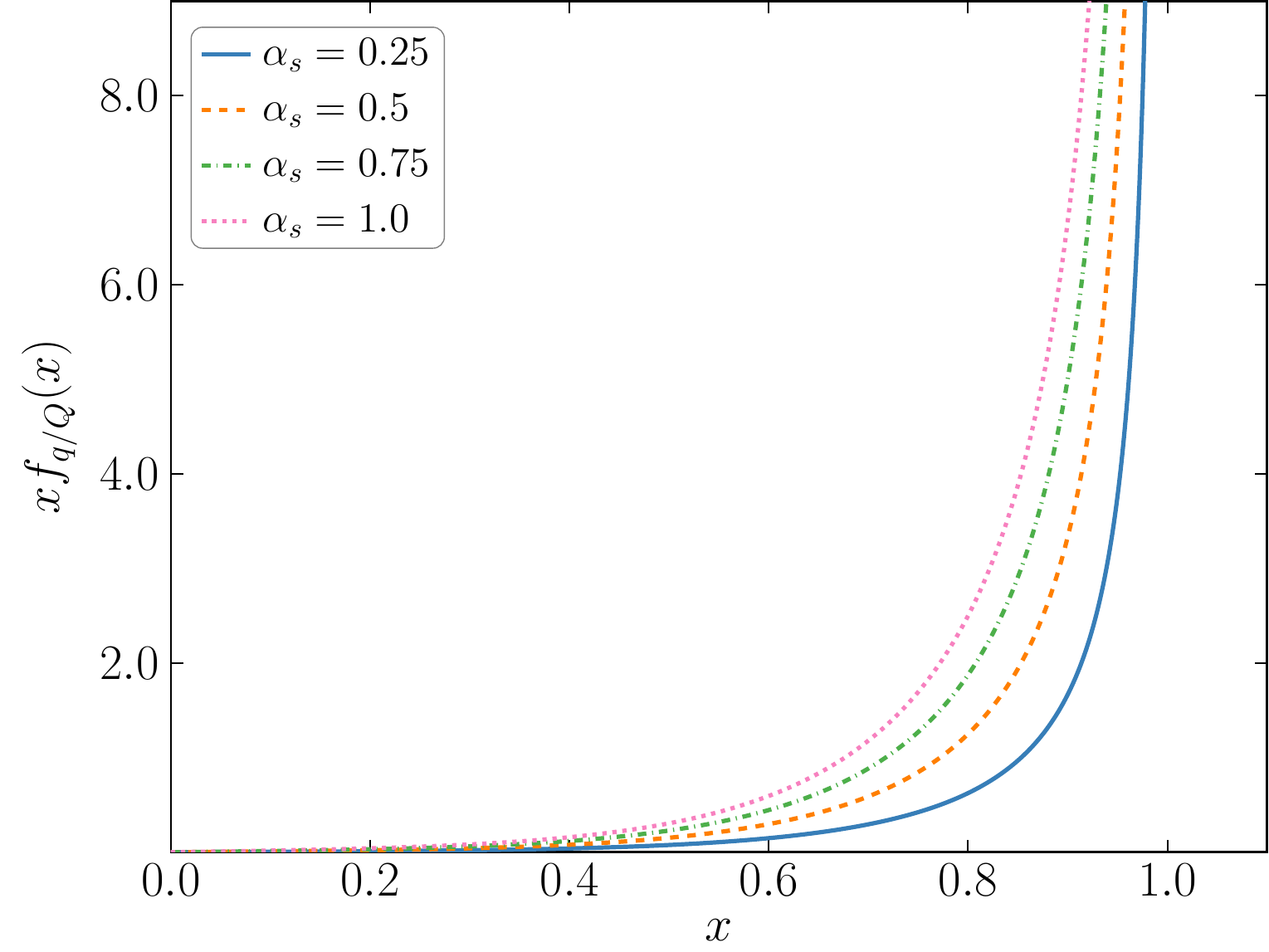}  \\[0.8em]
  \includegraphics[width=\columnwidth]{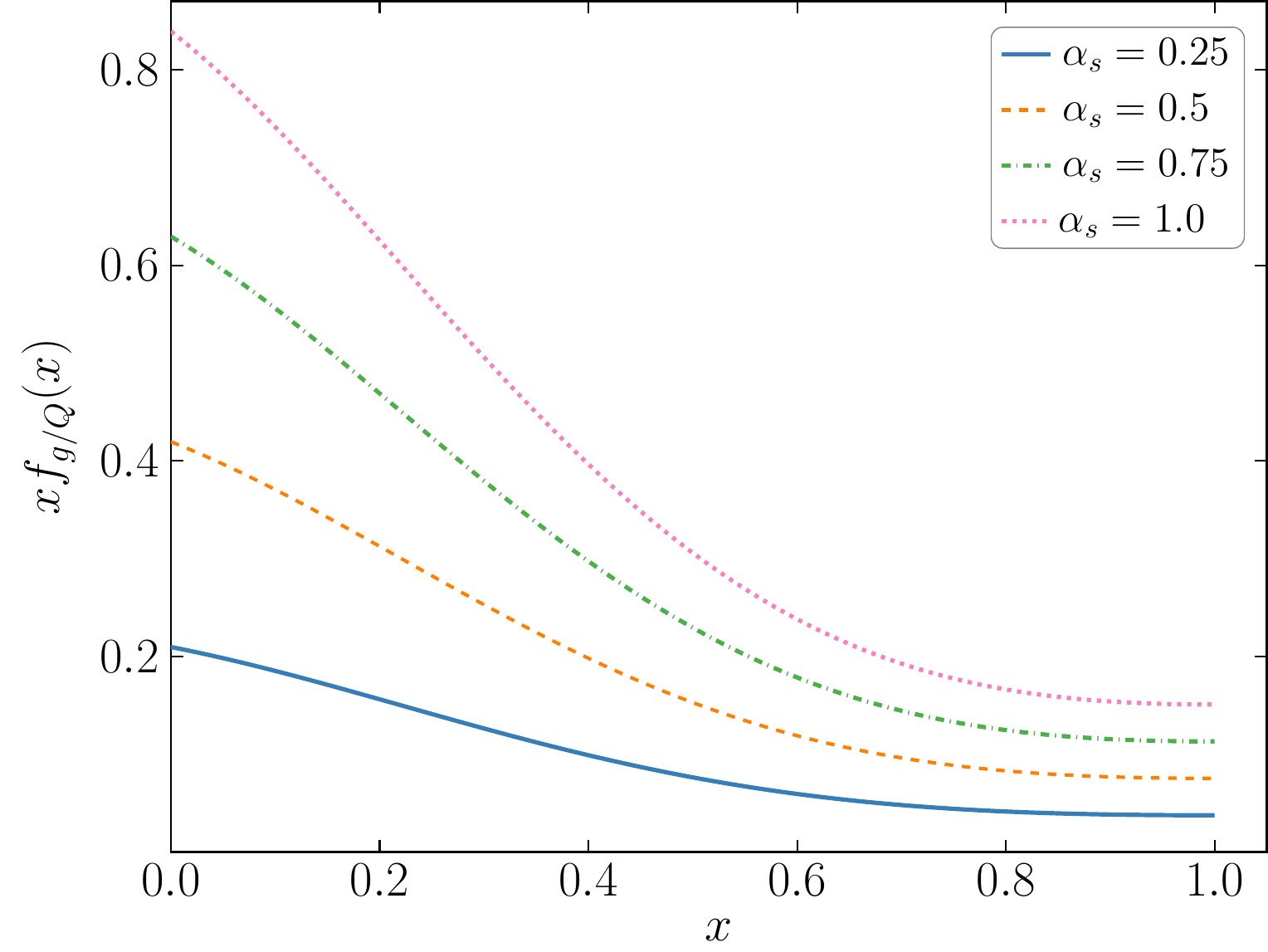} 
\caption{
  Results for the QTM quark and gluon PDFs for several values of $\alpha_s$ at fixed quark mass $M_q=0.4\,$GeV.
  The quark PDF notably has delta function contributions at $x=1$ that are not shown in this plot,
  and these delta function contributions are responsible
  for the negative Mellin moments in Tab.~\ref{Tab:Moments}.
  }
  \label{fig:PDFs-alpha}
\end{figure}

\begin{figure}[tbp]
  \centering
  \includegraphics[width=\columnwidth]{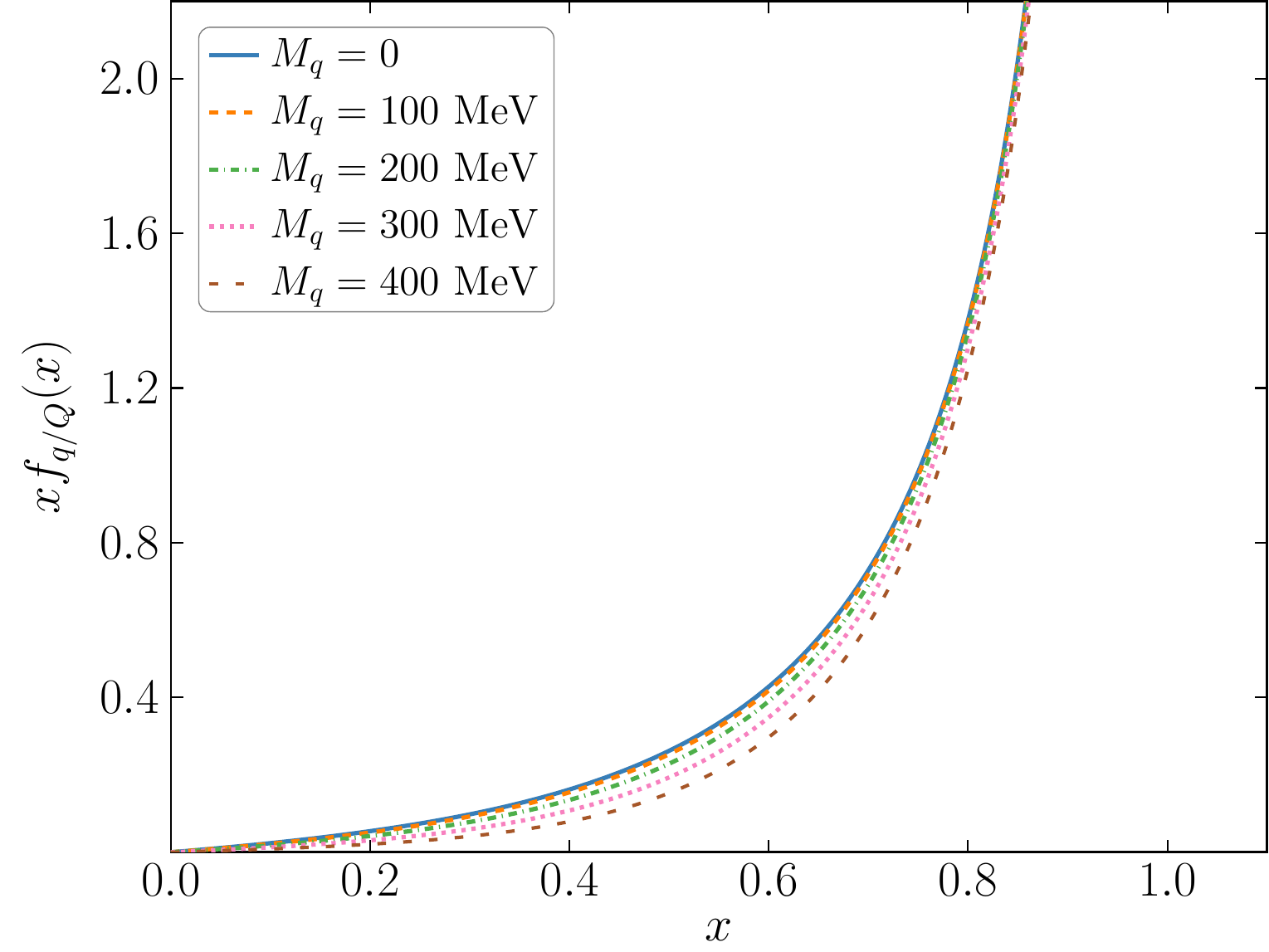} \\[0.8em]
  \includegraphics[width=\columnwidth]{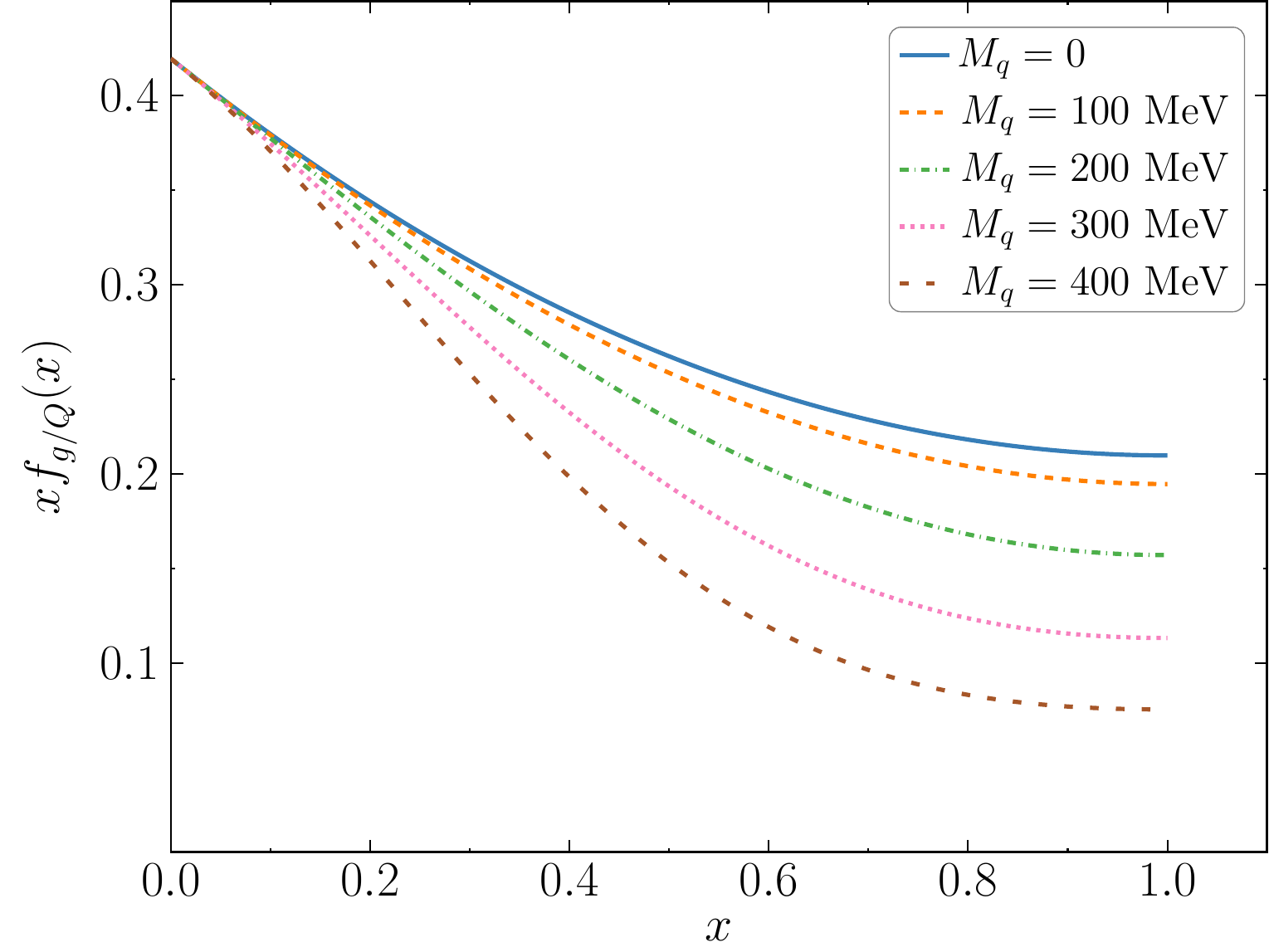}
\caption{Results for the QTM quark and gluon PDFs for several values of $M_q$ at fixed coupling strength $\alpha_s=0.5$. Note, the quark PDF also has delta function contributions at $x=1$ which are not shown.
  }
  \label{fig:PDFs-Mq}
\end{figure}
%===============================================================================

%%%%%%%%%%%%%%%%%%%%%%%%%%%%%%%%%%%%%%%%%%%%%%%%%%%%%%%%%%%%%%%%%%%%%%%%%%%%%%%%
\subsection{Numerical Results for the QTM}
The QTM PDFs contain IR and UV divergences that must be regularized, and therefore the numerical results are regularization scheme dependent. We employ proper time regularization with both UV and IR regulators~\cite{Ebert:1996vx,Hellstern:1997nv,Cloet:2014rja} to obtain the numerical results presented here, however, other schemes such as dimensional regularization (with MS or $\overline{\mathrm{MS}}$ subtraction of the $\epsilon^{-1}$ divergences) or implicit regularization can be used instead. (See App.~\ref{sec:integrals} for a dictionary to translate the formulas in this work into different regularization schemes.) For these numerical results we use $\Lambda_{\mathrm{IR}}=240\,$MeV and $\Lambda_{\mathrm{UV}}=645\,$MeV which are standard values~\cite{Cloet:2014rja}.

Fig.~\ref{fig:PDFs-alpha} gives results for the quark and gluon PDFs at different values of the coupling strength $\alpha_s$. Note, the $\delta(1-x)$ terms cannot be plotted. The gluon PDF is positive-definite and, aside from the $\delta(1-x)$ term from the Wilson line, so is the quark PDF. Since the PDFs are proportional to $\alpha_s$, we observe that the curves within each panel simply differ by an overall factor. At large $x$ the quark PDF is dominated by the $1/(1-x)$ term coming from the Wilson line in covariant gauges or the triangle diagram in light cone gauge. This behavior is different from familar quark PDFs inside hadrons and reflects that this quantity is a quark PDF inside itself.

The dependence of the PDFs on the physical quark mass $M_q$ is shown in Fig.~\ref{fig:PDFs-Mq}. The quark PDF only has slight dependence on the quark mass, with the non-delta function contributions being suppressed at larger quark masses. The gluon PDF is also suppressed for large quark masses and therefore gluons carry less momentum for larger quark masses.

%-----------------------------------------------------------
\begin{table}[tbp]
\addtolength{\tabcolsep}{7.5pt}
\addtolength{\extrarowheight}{1.5pt}
\caption{Results for the $s$ Mellin moments $\langle x^{s-1} \rangle = \int_{0}^1 dx\, x^{s-1}f(x)$ of the quark and gluon PDFs. In the second and third columns the quark PDF is split according to Eq.~\eqref{eqn:fqQ}. The fourth and fifth columns give the total quark and gluon moments. For these results the quark mass is set to $M_q = 0.4\,$GeV and the coupling strength to $\alpha_s=0.5$, which gives $Z_2 = 0.9933$ in a general covariant gauge at one loop.}
\label{Tab:Moments}
\begin{tabular}{c c c c c} \toprule
$s$ &
        $Z_2+\langle x^{s-1} \rangle_q^{\mathrm{tri}}$ &
        $\langle x^{s-1} \rangle_q^{W}$ &
        $\langle x^{s-1}  \rangle_q$ &
        $\langle x^{s-1}  \rangle_g$
        \\ \midrule
        1  &  1       &   ~~\,0     &  ~~\,1       &   --     \\
        2  &  0.9867  &  $-$0.1775  &  ~~\,0.8093  &  0.1907 \\
        3  &  0.9866  &  $-$0.3034  &  ~~\,0.6832  &  0.0647 \\
        4  &  0.9877  &  $-$0.4011  &  ~~\,0.5867  &  0.0351 \\  
        5  &  0.9888  &  $-$0.4807  &  ~~\,0.5081  &  0.0235 \\
        6  &  0.9896  &  $-$0.5479  &  ~~\,0.4418  &  0.0176 \\
        7  &  0.9903  &  $-$0.6059  &  ~~\,0.3843  &  0.0140 \\
        8  &  0.9908  &  $-$0.6571  &  ~~\,0.3337  &  0.0118 \\ 
        9  &  0.9912  &  $-$0.7027  &  ~~\,0.2884  &  0.0101 \\
        10 &  0.9915  &  $-$0.7440  &  ~~\,0.2475  &  0.0088 \\
    \vdots &  \vdots  &   \vdots    &     \vdots &  \vdots \\
        18 &  0.9926  &  $-$0.9791  &   ~~\,0.0135 &  0.0045 \\  
        19 &  0.9927  &  $-$1.0010  &  $-$0.0084 &  0.0043 \\
        20 &  0.9927  &  $-$1.0219  &  $-$0.0292 &  0.0040 \\ 
        \bottomrule
\end{tabular}
\end{table}
%-----------------------------------------------------------

In Tab.~\ref{Tab:Moments} we present Mellin moments of the quark and gluon PDFs. To reveal the importance of the Wilson line to the moments, especially the momentum sum rule,  we split the quark PDF into two terms: one containing $Z_2$ and triangle diagram contributions, and the other contribution containing the Wilson line. In a general covariant gauge these results are shown in the second and third columns of Tab.~\ref{Tab:Moments}, with the fourth and fifth columns containing the total quark and gluon Mellin moments, respectively, which are both gauge independent.

The results in Tab.~\ref{Tab:Moments} show that the quark number sum rule is satisfied without needing to account for the Wilson line, a result proven earlier analytically. However, the momentum sum rule is oversaturated without the negative contribution from the Wilson line. With $M_q = 0.4\,$GeV and $\alpha_s= 0.5$ we find that the quarks carry about 81\% of the total light cone momentum and gluons carry about 19\%. The Wilson line contribution to the momentum sum rule is $-0.1775$, which is the same in all covariant gauges, because we are working at one-loop in the on-shell remormalization scheme. This result explicitly illustrates that Wilson line contributions can be large in covariant gauges. 

This has important implications for model calculations 
%(e.g. using the Dyson-Schwinger equations~\cite{}) 
in covariant gauges that do not explicitly include gauge link contributions to quantum correlation functions. For example, if one were to calculate only quark local operators, and infer the momentum fraction carried by gluons by imposing the momentum sum rule, then one must include Wilson line contributions in order to obtain the correct gluon momentum fraction. Had we not accounted for the Wilson line in this calculation, for instance, and had inferred gluon momentum by what was missing from the momentum sum rule, we would have underestimated the gluon momentum fraction by $0.1775$, that is, obtaining $\lf<x\rg>_g = 0.0133$ instead of the correct value of $\lf<x\rg>_g = 0.1907$. This clearly indicates the importance of accounting for the Wilson line contributions.

An interesting feature of the QTM is that the Mellin moments of the quark PDF become negative starting at $s=19$ for the parameters used in Tab.~\ref{Tab:Moments}. This occurs in covariant gauges specifically because of the Wilson line contribution, and in light cone gauge because of the extra term in the triangle diagram in Eq.~(\ref{eq:tri_LC}). This has a major implication for applying the QTM to calculating hadron PDFs. That is, if a standard convolution model is used in the impulse approximation, and if the PDFs for the hadron that come from a quark-only effective theory are positive definite (body PDFs), then incorporating the QTM PDFs will make the hadron's quark PDF become negative near $x \sim 1$. We illustrate this explicitly for the pion in Sec.~\ref{sec:pion}.

Tab.~\ref{Tab:Moments2} gives the second Mellin moment of the quark and gluon PDFs for various values of the coupling strength. The momentum carried by the gluons increases with the coupling strength, and its worth highlighting that the gluons contribute approximately $28\%$ of the quark target light cone momentum with $\alpha_s= 0.75$. This is approximately equal to the gluon momentum fraction found empirically for the pion in Ref.~\cite{Barry:2018ort}. Therefore, this simple one-loop QTM can give a physically reasonable value for the gluon momentum fraction for realistic values of $\alpha_s$.

%-----------------------------------------------------------
\begin{table}[tbp]
\addtolength{\tabcolsep}{9.3pt}
\addtolength{\extrarowheight}{1.5pt}
\caption{Results for the second Mellin moment $\langle x \rangle = \int_{0}^1 dx \, xf(x)$ of the quark target model PDFs for several values of the coupling strength $\alpha_s$, given at a fixed quark mass $M_{q}=0.4~\mathrm{GeV}$. In the second and third columns the quark PDF is split according to Eq.~\eqref{eqn:fqQ}. The fourth and fifth columns give the total quark and gluon moments.}
    \label{Tab:Moments2}
    \begin{tabular}{c c c c c} \toprule
      $\alpha_{s} $ &
      $Z_2+\langle x \rangle_q^{\mathrm{tri}}$ &
      $\langle x \rangle_q^{W}$ &
      $\langle x \rangle_q$ &
      $\langle x \rangle_g$
      \\ \midrule
      0.25 &  0.9934   &  $-$0.0887  &  0.9046  &  0.0954 \\
      0.50 &  0.9867   &  $-$0.1775  &  0.8093  &  0.1907 \\
      0.75 &  0.9800   &  $-$0.2662  &  0.7139  &  0.2861 \\
      1.00 &  0.9735   &  $-$0.3549  &  0.6185  &  0.3815 \\
      \bottomrule
    \end{tabular}
  \end{table}
%-----------------------------------------------------------

%%%%%%%%%%%%%%%%%%%%%%%%%%%%%%%%%%%%%%%%%%%%%%%%%%%%%%%%%%%%%%%%%%%%%%%%%%%%%%%%
\section{QUARK TARGET MODEL WITH A GLUON MASS\label{sec:gmass}}
In this section we introduce a gluon mass in order to study its impact on the QTM PDFs. Results from both lattice QCD~\cite{Leinweber:1998uu,Oliveira:2010xc,Duarte:2016iko} and Dyson-Schwinger equations~\cite{Bhagwat:2003vw,Bhagwat:2006tu,Aguilar:2011ux} suggest that the gluon in QCD may acquire an dynamically generated effective mass at low momentum. The mass is apparently generated by the full structure of QCD, including gluon self-interactions to all orders in $\alpha_s$. It therefore isn't possible to reproduce dynamical gluon mass generation in an effective model of gluons at any finite order in $\alpha_s$. Nonetheless, the effect of a gluon mass can be incorporated in a model by placing an explicit gluon mass term in the Lagrangian and it has been shown to provide good agreement with lattice results for the gluon and ghost propagators already at one-loop~\cite{Tissier:2010ts,DallOlio:2020xpu}. Therefore, it is worthwhile to explore the QTM PDFs using a finite gluon mass.
\subsection{Formalism for a Gauge Invariant Gluon Mass}
A naive Lagrangian mass term for gluons reads~\cite{Curci:1976bt,Tissier:2010ts}:
\begin{align}
  \mathcal{L}_{\mathrm{mass}}^{(\mathrm{naive})}
  =
  m_g^2\, \mathrm{Tr} \left[
    A_\mu(x) A^\mu(x)
    \right],
  \label{eqn:L:naive}
\end{align}
which has been widely used to describe lattice QCD results on gluon and ghost two-point functions~\cite{Tissier:2010ts,DallOlio:2020xpu} but this term notoriously violates gauge invariance. In addition, as we shall see explicitly, including this mass term in QTM PDF calculations violates the momentum sum rule in any gauge. This occurs because a gauge transformation can be used to introduce explicit dependence on spacetime coordinates $x$ into the Lagrangian, nullifying conservation of the energy-momentum tensor.

A gauge-invariant Lagrangian mass term for gluons has been suggested by Cornwall~\cite{Cornwall:1981zr}:
\begin{align}
\hs*{-1.6mm}  \mathcal{L}_{\mathrm{mass}}
  =
  m_g^2\, \mathrm{Tr} \left[
    \left(
    A_\mu(x) - \frac{1}{ig}\Big(\partial_\mu V(\theta(x))\Big)V^{-1}(\theta(x))
    \right)^2
    \right],
  \label{eqn:L:mass}
\end{align}
where the field $V(\theta(x)) = \exp\{i\theta(x)\}$ transforms under the fundamental representation of the color group. This mass term makes the theory non-renormalizable, which would render the model inapplicable beyond one-loop, but since the calculations herein are done at one loop this is not an issue for this work.

Including the mass term of Eq.~(\ref{eqn:L:mass}) has primarily two effects on the calculation of PDFs in the QTM. The first is that a gluon mass now appears in the gluon propagator, which for covariant gauges and light cone gauge take the form
\begin{subequations}
  \label{eqn:Dmn}
  \begin{align}
    \label{eqn:Dmn:covariant}
    D^{\mathrm{cov}}_{\mu\nu}(k) 
    &=
    \frac{-1}{k^2-m_g^2+i0} \left[
      g_{\mu\nu}
      -
      (1-\xi)\, \frac{ k_\mu k_\nu }{ k^2 -\xi m_g^2 }
      \right],
    \\
    D^{\mathrm{LC}}_{\mu\nu}(k) 
    &=
    \frac{-1}{k^2-m_g^2+i0} \left[
      g_{\mu\nu} - \frac{k_\mu n_\nu + n_\mu k_\nu}{(kn)}
      \right].
  \end{align}
\end{subequations}
The gluon propagators given in Eqs.~\eqref{eqn:Dmn} take the same form if either the naive gluon mass term of Eq.~(\ref{eqn:L:naive}) or the Cornwall mass term of Eq.~(\ref{eqn:L:mass}) is added to the QTM Lagrangian. Nevertheless, in general the covariant gauge result violates the QCD Slavnov-Taylor identity for the gluon propagator~\cite{Slavnov:1972fg,Taylor:1971ff}, which states $p_\mu\,D^{\mu\nu}(p) = \xi\  p^\nu/p^2$, and follows from BRST symmetry. It implies that only the transverse piece of the gluon propagator is dressed in covariant gauges. Coincidentally, in the Landau gauge ($\xi=0$) the gluon propagator does remain transverse, even with an explicit gluon mass term in the Lagrangian.

Including a gluon mass in the gluon propagators, as in Eqs.~\eqref{eqn:Dmn}, and calculating the diagrams in Figs.~\ref{fig:qtm_pdf_q} and \ref{fig:qtm_g_gluon} gives the following gauge invariant results for the quark and (naive) gluon PDFs:
\begin{align}
\label{eq:fqm}
f_{q/Q}(x;m_g) &= f_{q/Q}(x)  \no \\
&\hs*{10mm}
- 4\,g^2 C_F\,m_g^2 \, x(1-x)\,I_3\,\big(w_q(x)\big), \\
\label{eq:fgm}
x\,f^{\mathrm{naive}}_{g/Q}(x;m_g) &= x\,f_{g/Q}(x) \no \\
&\hs*{-11.5mm}
- 4\,g^2 C_F\,m_g^2\,\Big[x^2(1-x) + 2\,(1-x)^2\Big]\,I_3(w_g(x)),
\end{align}
where in these expressions $w_q(x) = (1-x)^2 M_q^2 + x\,m_g^2$ and $w_g(x) = w_q(1-x)$. The functions $f_{q/Q}(x)$ and $f_{g/Q}(x)$ are our earlier results given in Eqs.~\eqref{eqn:fqQ} and \eqref{eqn:fgQ}, except with this new Feynman mass parameter that includes a gluon mass in the basic integrals. The quark field renormalization constant, $Z_2$ from Eqs.~\eqref{eq:Z2cov} and \eqref{eq:Z2LC}, is also updated accordingly.

The quark PDF given in Eq.~\eqref{eq:fqm}, together with the similarly modified $Z_2$, exactly satisfies the quark number sum rule for any $m_g$. However, the quark and gluon PDF results of Eqs.~\eqref{eq:fqm} and \eqref{eq:fgm} do not satisfy the  momentum sum rule for any finite $m_g$. This can easily be seen because the symmetry $f_{q/Q}(x;m_g) = f^{\mathrm{naive}}_{g/Q}(1-x;m_g)$ is violated by the last term in Eq.~\eqref{eq:fgm}, which is proportional to $m_g^2$. As such, using the naive gluon mass term of Eq.~\eqref{eqn:L:naive} not only violates gauge invariance but also momentum conversation as expressed via the momentum sum rule. This is the reason why the superscript ``naive'' is applied to the massive gluon PDF of Eq.~\eqref{eq:fgm}.

%If the change in the gluon propagator were the only effect of introducing gluon mass\emdash as is the case when the naive Lagrangian of Eq.~(\ref{eqn:L:naive}) is used\emdash then we would observe violation of the momentum sum rule. However, the Cornwall Lagrangian mass term saves the momentum sum rule, as we shall presently see.

%which would have guaranteed the momentum sum rule if this were the \emph{only} contribution to the gluon PDF.  

% because this PDF now has additional contributions from the field $\theta(x)$. This is the second major impact of the Cornwall mass term. Crucially, we do not observe the symmetry $f_{q/Q}(x;m_g) = f^{\mathrm{tri}}_{g/Q}(1-x;m_g)$, which would have guaranteed the momentum sum rule if this were the \emph{only} contribution to the gluon PDF. I
% nstead, the symmetry is violated by an amount
% %
% \begin{align}
% &f_{q/Q}(x;m_g) - f^{\mathrm{tri}}_{g/Q}(1-x;m_g) \no \\
% %
% &\hs*{30mm}
% =
% 8\,g^2 C_F\,m_g^2\, \frac{(1-x)^2}{x} \, I_3\big(W_g(x)\big),
% \end{align}
% %
% which is proportional to $m_g^2$. 

The Cornwall mass term of Eq.~\eqref{eqn:L:mass} not only introduces an explicit gluon mass ($m_g$) but also the additional auxiliary field $\theta(x)$. As we shall see, this theta-field also has a PDF which will restore the momentum sum rule. We find that the bilocal light cone correlator defining the theta-field PDF is
\begin{align}
x\,f_{\theta/Q}(x) &= -\frac{2\,m_g^2}{g^2}\ \frac{n^\mu n^\nu}{P\cdot n} 
\int \frac{\mathrm{d}\lambda}{2\pi}\, e^{ix\lambda\,P\cdot n} \no \\
&\hs*{-5mm}
  \Big< Q(P) \Big|
  \left[ D_\mu V\big(\theta(0)\big) \right]
  V^{-1}\big(\theta(0)\big)\,
  W_A(0,n\lambda) \no \\
&\hs*{11mm}
\times
  \left[ D_\nu V\big(\theta(n\lambda)\big) \right]
  V^{-1}\big(\theta(n\lambda)\big)
  \Big| Q(P)\Big>.
\end{align}
Explicit calculation gives the following theta-field PDF:
\begin{align}
x\,f_{\theta/Q}(x) =
  8\,g^2\, C_F \, m_g^2\, (1-x)^2 \, I_3\big(w_g(x)\big).
\end{align}
Since the theta-field is interpreted as an instanton in the gluon field~\cite{Cornwall:1981zr}, $f_{\theta/Q}(x)$ should be considered a contribution to the gluon PDF. Therefore, the full gluon PDF with the Cornwall mass term is given by the sum of $f^{\mathrm{naive}}_{g/Q}(x;m_g)$ and $f_{\theta/Q}(x;m_g)$, and reads
\begin{align}
\hs*{-1mm}
f_{g/Q}(x;m_g) %&= f^{\mathrm{naive}}_{g/Q}(x;m_g) + f_{\theta/Q}(x;m_g), \no \\
%
%&\hs*{-2mm}
= f_{g/Q}(x) - 4\,g^2 C_F\,m_g^2\,x\,(1-x)\,I_3\big(w_g(x)\big).
\end{align}
This expression satisfies $f_{q/Q}(x;m_g) = f_{g/Q}(1-x;m_g)$ which guarantees the momentum sum rule as the quark number sum rule is satisfied. Therefore, the theta-field contribution to the gluon PDF must be accounted for to ensure observance of the momentum sum rule. Further, since $f_{\theta/Q}(x;m_g)$ is gauge invariant there is no choice of gauge that can restore the momentum sum rule when only the naive gluon mass term is considered.

%-----------------------------------------------------------
\begin{figure}[tbp]
  \centering
  \includegraphics[width=\columnwidth]{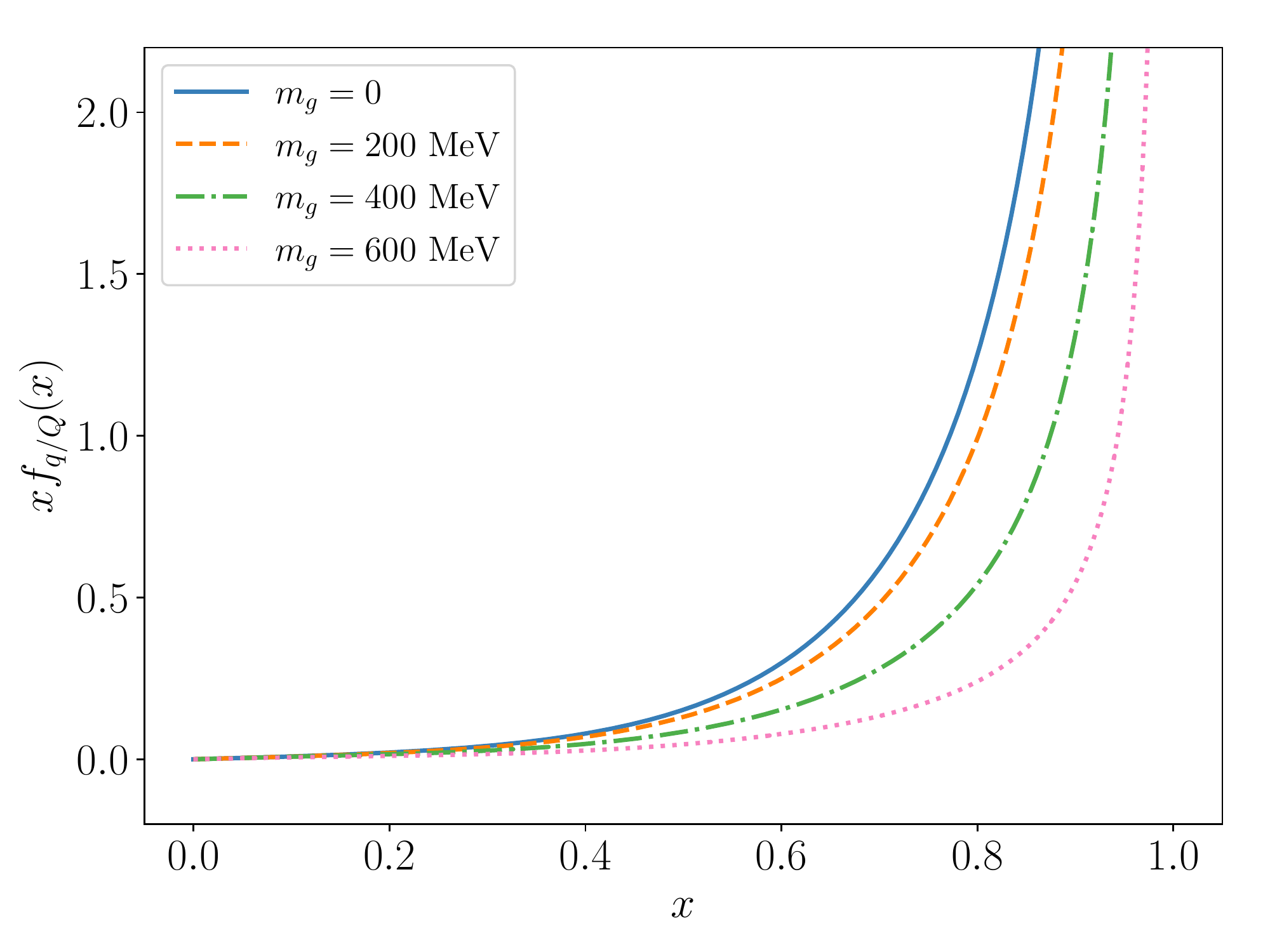}
  \includegraphics[width=\columnwidth]{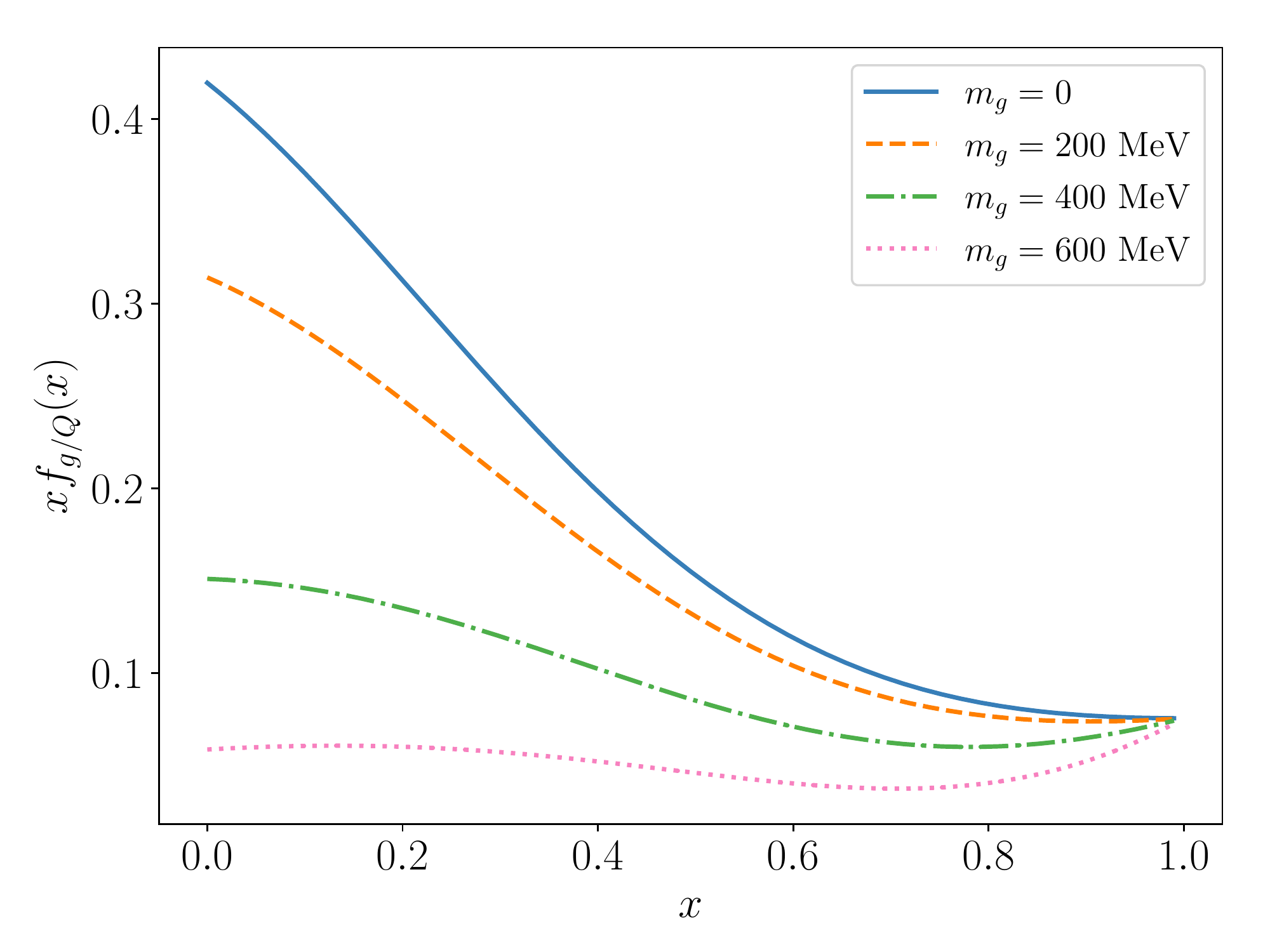}
\caption{Quark and gluon PDFs in the QTM for several values of the Cornwall gluon mass $m_g$, with the gluon PDF defined to include the theta-field contribution. The plotted quark PDF also includes $\delta(1-x)$ contributions, and $\alpha_s = 0.5$ and $M_q=0.4\,$GeV were used in these calculations.}
\label{fig:pdf:mg}
\end{figure}
%-----------------------------------------------------------

%===============================================================================
%===============================================================================
\subsection{Numerical Results with Gluon Mass}
Results for the QTM quark and gluon PDFs, for several values of the Cornwall gluon mass, are presented in Fig.~\ref{fig:pdf:mg}. The plotted quark PDF does not show the $\delta(1-x)$ terms which contribute to the quark number and momentum sum rules. The introduction of a gluon mass suppresses both PDFs, with the quark number and momentum sum rules preserved specifically
by an increase in the $\delta(1-x)$ terms.
Physically, this occurs because it costs more to radiate a gluon when the gluon has mass.
Therefore, with all other parameters fixed, a larger gluon mass results
in gluons carrying a smaller fraction of the quark target's light cone momentum.
On the other hand, once a gluon is radiated,
it is more likely to carry a large fraction of light cone momentum owing to its mass.
This has the effect of reshaping the gluon PDF, so that as $m_g$ increases a larger fraction of the support for the gluon PDF is at large $x$. This can be observed in the shapes of the curves in the lower panel
of Fig.~\ref{fig:pdf:mg}.

%-----------------------------------------------------------
\begin{figure}[tbp]
\centering
\includegraphics[width=\columnwidth]{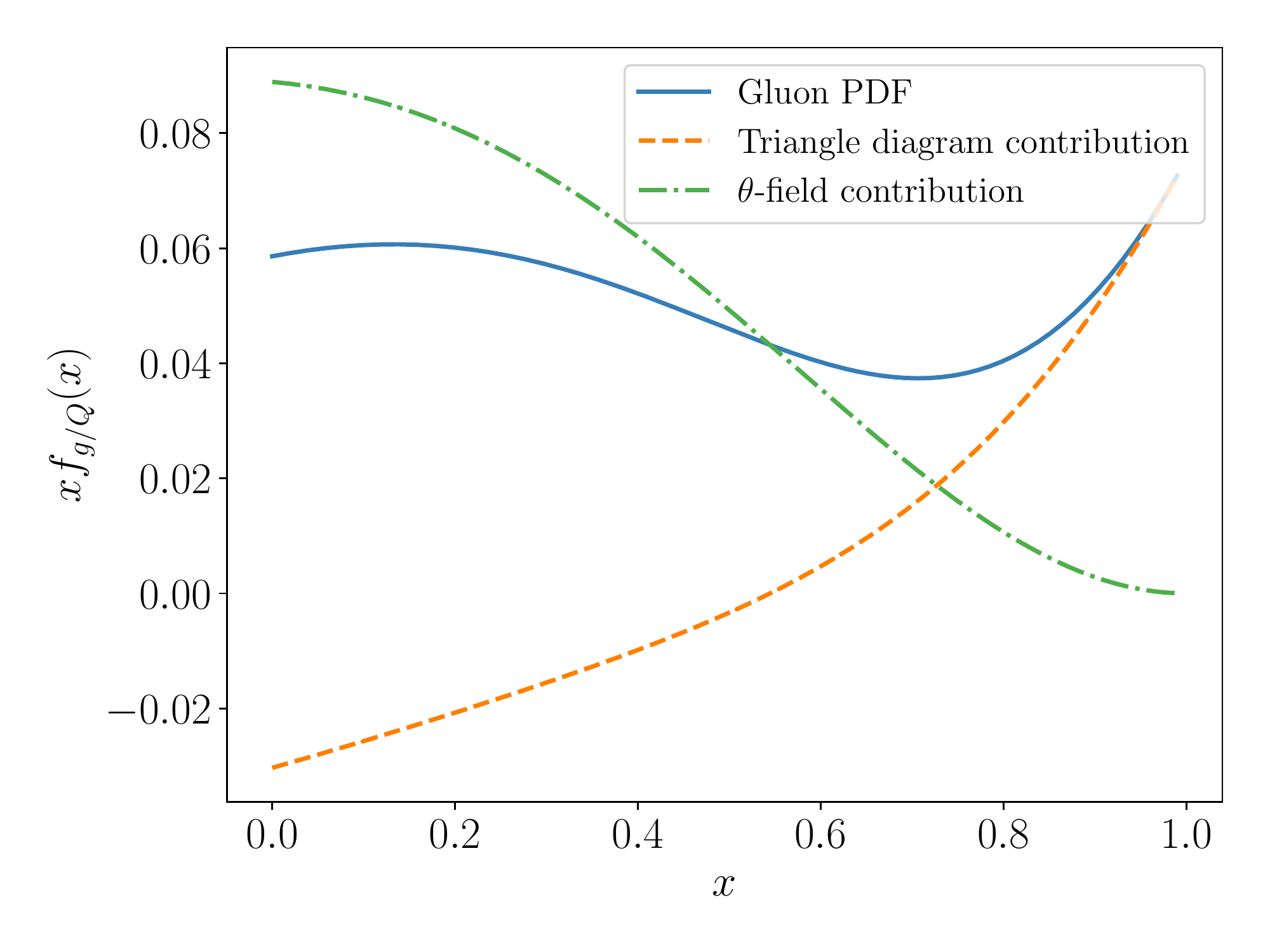}
\caption{Gluon PDFs in the quark target model, separated into naive and theta-field contributions. We use $\alpha_s=0.5$, $M_q = 0.4\,$GeV, and $m_g = 0.6\,$GeV.}
\label{fig:pdf:mg:breakdown}
\end{figure}
%-----------------------------------------------------------

In Fig.~\ref{fig:pdf:mg:breakdown}, we separate the gluon PDF into naive and theta-field contributions, where $m_g=0.6\,$GeV is used for the gluon mass to illustrate an extreme case, and we take  $\alpha_s=0.5$ and $M_q = 0.4\,$GeV. In general, the naive diagram makes more significant positive contributions at large $x$, while the theta-field dominates at small $x$.
For this large value of the gluon mass, the naive contribution goes negative at small $x$, however, the sum of both contributions remains positive-definite.

%-----------------------------------------------------------
\begin{table}[bp]
\addtolength{\tabcolsep}{2.8pt}
\addtolength{\extrarowheight}{1.5pt}
\caption{Contributions to the quark and gluon momentum fractions for various values of the Cornwall gluon mass (in GeV). The quark momentum fractions are separated according to Eq.~\eqref{eqn:fqQ} and the gluon momentum fractions are separated into naive and theta-field contributions. The other parameters are $\alpha_s=0.5$ and $M_q= 0.4\,$GeV.}
  \label{tab:mg:xg}
  \begin{tabular}{@{}cccccccc@{}}
    \toprule
    $m_g$ &
    $Z_2+\langle x \rangle_q^{\mathrm{tri}}$ &
    $\langle x \rangle_q^{W}$ &
    $\langle x  \rangle_q$ &
    $\langle x \rangle_g^{\mathrm{naive}} $ &
    $\langle x \rangle_\theta $ &
    $\langle x \rangle_g $ \\
    \hline
    $0$       &
    $0.9867$  &
   $-0.1775$  &
    $0.8093$  & 
    $0.1907$  &
    $0$       &
    $0.1907$
    \\
    $0.2$     &
    $0.9885$  &
   $-0.1454$  &
    $0.8431$  & 
    $0.1293$  &
    $0.0276$  &
    $0.1569$
    \\
    $0.4$     &
    $0.9920$  &
   $-0.0878$  &
    $0.9042$  & 
    $0.0434$  &
    $0.0524$  &
    $0.0958$
    \\
    $0.6$     &
    $0.9951$  &
   $-0.0462$  &
    $0.9489$  & 
    $0.0045$  &
    $0.0466$  &
    $0.0511$
    \\
    \bottomrule
  \end{tabular}
\end{table}
%-----------------------------------------------------------

The relative importance of the theta-field contribution to the total gluon PDF and momentum fraction are presented in Tab.~\ref{tab:mg:xg}. When the gluon is massive, a significant portion of the gluon's light cone momentum is contained in the theta-field PDF, rather than in the traditional gluon PDF. In fact, for $m_g=0.6\,$GeV, the theta-field carries about $90\%$ of the gluon's light cone momentum, however, as $m_g$ become large both the naive and theta-field contribution tend to zero and the quark PDF approaches $\delta(1-x)$.

Through these numerical examples we observe that the auxiliary theta-field is essential to accommodate massive gluons, since it carries a significant portion of the gluons' light cone momentum and because it is necessary to guarantee that the gluon PDF be positive-definite.

%%%%%%%%%%%%%%%%%%%%%%%%%%%%%%%%%%%%%%%%%%%%%%%%%%%%%%%%%%%%%%%%%%%%%%%%%%%%%%%%
\section{PION PDFs\label{sec:pion}}
In this section, we combine the QTM results for the quark and gluon PDFs for a dressed quark, with existing results for hadron PDFs from low-energy effective models with only quark degrees of freedom. We consider the pion as a concrete example, because of its prominent role in QCD as the Nambu-Goldstone boson associated with dynamical chiral symmetry breaking. In addition, both experimental data~\cite{Conway:1989fs} and phenomenological PDF parametrizations~\cite{Barry:2018ort} are available.

The QTM-modified pion PDFs are obtained by taking a convolution of the pion PDF obtained from the effective model, $f_{Q/\pi}(x)$, which only includes dressed quark degrees of freedom, with the QTM PDFs, $f_{q,g/Q}(x)$, which include quark and gluon distributions inside the dressed quarks. That is, the quark target is identified with the effective theory's quark degrees of freedom, both denoted by $Q$. The complete quark and gluon PDFs of the pion are then given by the convolution:
\begin{align}
%f_{q,g/\pi}(x) &= \sum_Q
%  \int_x^1 \frac{\mathrm{d}y}{y}\, f_{q,g/Q}(y)\, f_{Q/\pi}\left(\frac{x}{y}\right).
f_{q,g/\pi}(x) &= \sum_Q
  \iint_0^1 \mathrm{d}z\, \mathrm{d}y\, \delta(x-yz)\, f_{q,g/Q}(y)\, f_{Q/\pi}(z).
  \label{eqn:convolution}
\end{align}
Since the QTM PDFs satisfy the quark number and momentum sum rules for the dressed quark target, this convolution guarantees that these fundamental sum rules are also satisfied for the total pion PDFs,  $f_{q,g/\pi}(x)$, provided the pion ``body PDF'' $f_{Q/\pi}(x)$ from the effective model also obeys these sum rules.

As a concrete example, we consider the quark PDF of the $\pi^+$ in the NJL model, which is calculated via:
\begin{align}
f_{Q/\pi}(x)
  &= -\frac{Z_{\pi}}{4\,P\cdot n}
  \int \frac{\mathrm{d}^4k}{(2\pi)^4}\
  \delta\left(x-\frac{k\cdot n}{P\cdot n}\right)   \no \\
&\hs*{-5mm}\times
  \mathrm{Tr}\big[\gamma_5\, \tau_-\, iS(k)\,
  \slashed{n}\,(1\pm \tau_3)\,iS(k)\,\gamma_5\,\tau_+\,iS(k-p) \big],
\end{align}
where the trace is over color, flavor and Dirac indices. The result for $Q=U,\bar{D}$ is found to be~\cite{Hutauruk:2018zfk}:
\begin{equation}
  f_{Q/\pi}(x)= 3\,Z_\pi\left[8\,x(1-x)\,m_\pi^2 - 4\,I_2(w_\pi(x))\right],
\end{equation}
where $w_\pi(x) = M_q^2-x(1-x)m_\pi^2$, $m_\pi$ is the pion mass, and $Z_\pi$ can be calculated directly or obtained from the baryon number sum rule.

To fully define the model, it is necessary to determine a value for $\alpha_s$ to use both in the QTM calculations and in the DGLAP evolution equations to connect the low-energy effective theory to large-$Q^2$ experiments and phenomenology. It is also necessary to pick a value for $m_g$. We consider both the cases of zero gluon mass, $m_g=0$, and a finite Cornwall mass, $m_g=0.4\,$GeV. The model scale is determined by requiring that the pion's gluon momentum fraction content match that found by the JAM analysis of Ref.~\cite{Barry:2018ort} after NLO DGLAP evolution, where at the charm threshold is around 30\%. For the $m_g=0$ case this gives $\alpha_{s0}=0.579$ and, using the NLO equation for $\alpha_s(Q^2)$, a model scale of $Q_0^2=0.82\,$GeV$^2$. Similarly, for $m_g=0.4\,$GeV we find a model scale of $Q_0^2=0.58\,$GeV$^2$. It is worth mentioning that, already at the model scale for $m_g=0$, gluons carry $22\%$ of pion's light cone momentum and the model scale is much larger than typically used in models with only dressed quarks~\cite{Cloet:2007em,Ninomiya:2017ggn}. With $m_g=0.4\,$GeV the gluon light cone momentum fraction is $7\%$ at the corresponding model scale.

\begin{figure}[tbp]
  \centering
  \includegraphics[width=\columnwidth]{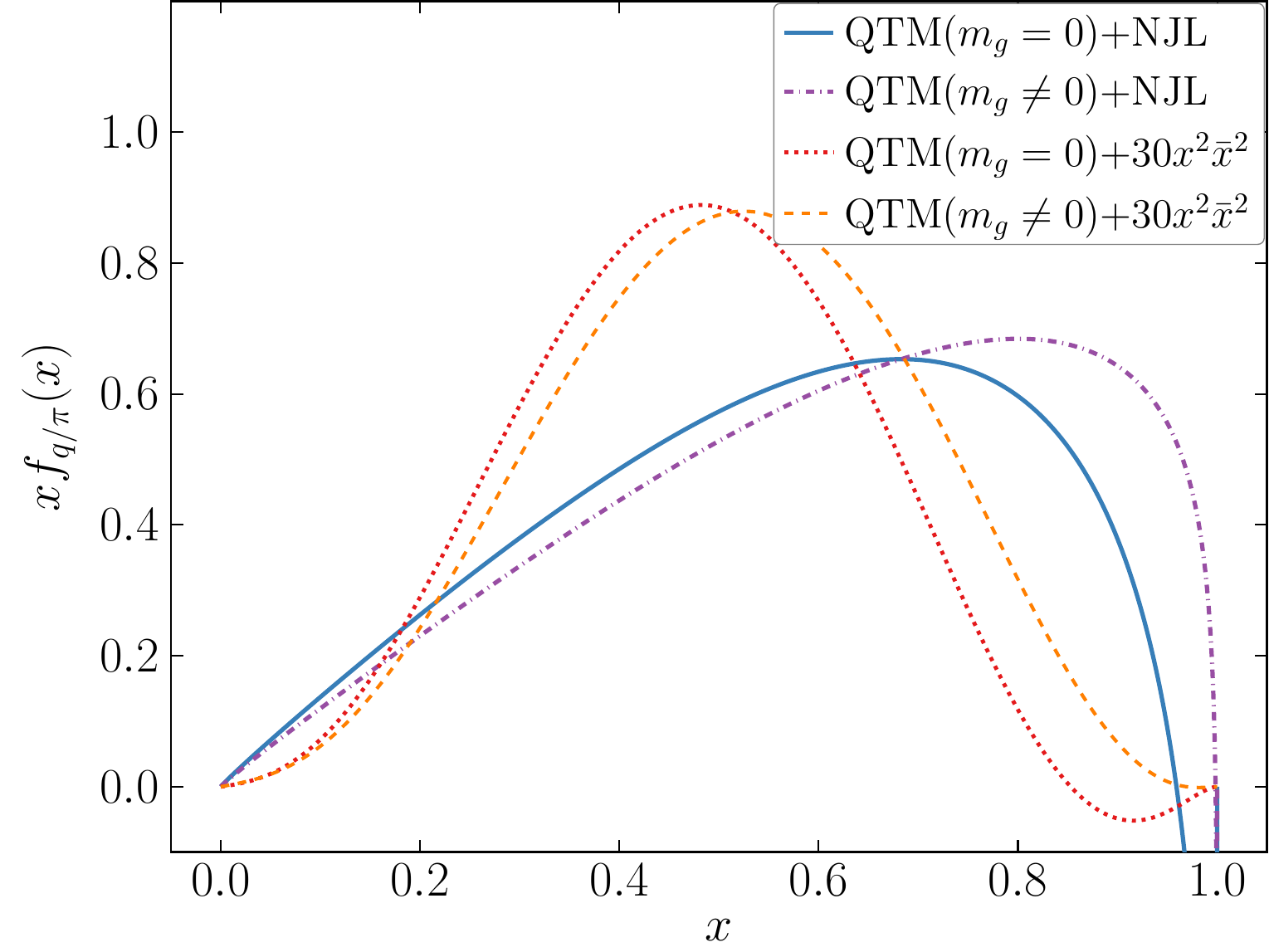}  \\[0.8em]
  \includegraphics[width=\columnwidth]{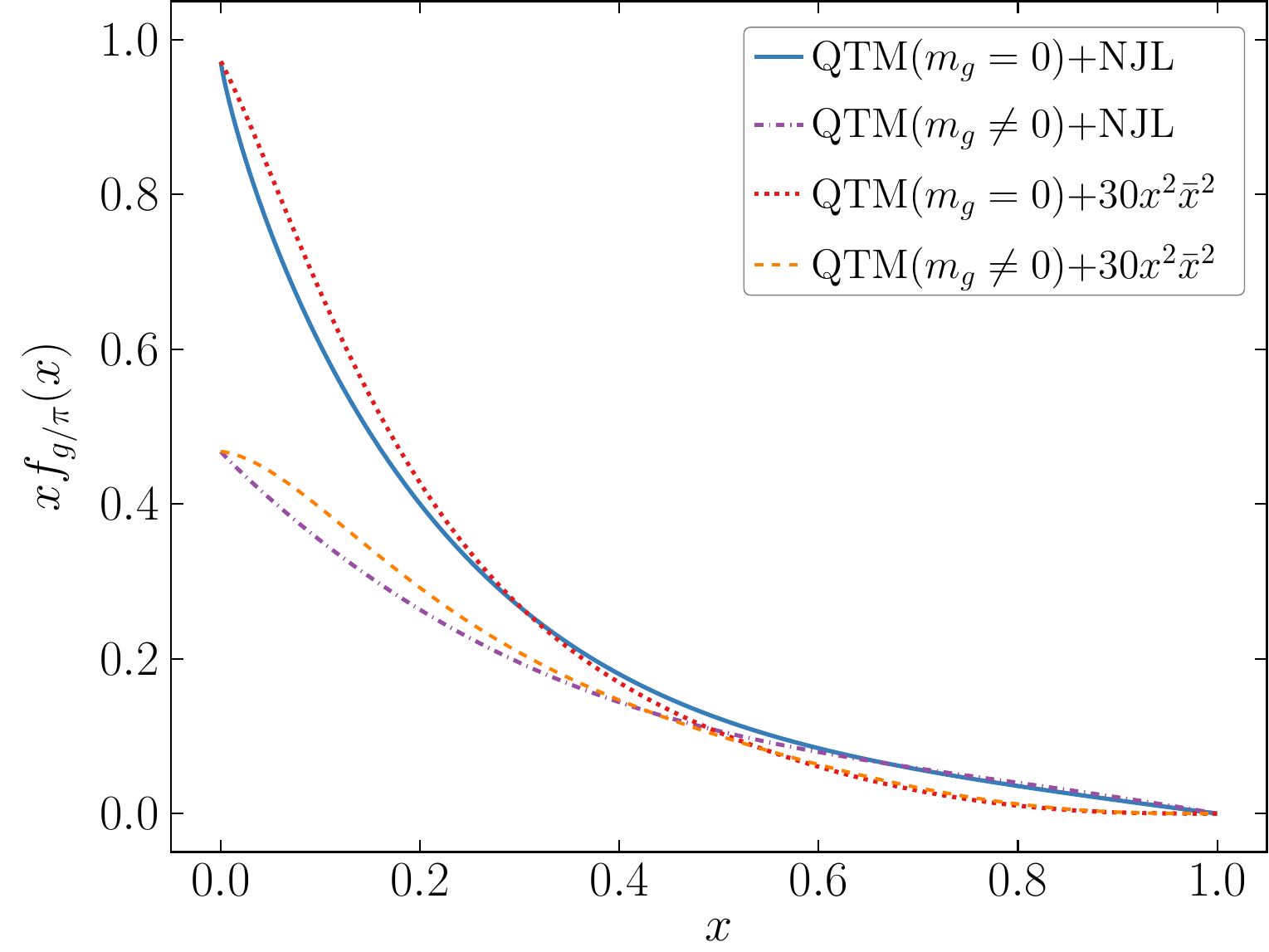}
  \caption{Quark (top panel) and gluon (bottom panel) PDFs of the pion at the model scales. For the body PDFs we use the NJL model result and the simple parameterization $30\,x^2\bar{x}^2$, where $\bar{x}=1-x$. For the QTM we show results for $m_g=0$ and $m_g=0.4\,$GeV, which correspond to model scales of $Q_0^2 = 0.82\,$GeV$^2$ and $Q_0^2 = 0.58\,$GeV$^2$, respectively.}
\label{fig:pdf:xfpiQ0}
\end{figure}

To illustrate how the PDF results depend on the underlying model body PDFs, we also consider a naive pion body PDF of $f_{Q/\pi}(x)=30x^2(1-x)^2$ as a reference. The results using both body PDFs are presented in Fig.~\ref{fig:pdf:xfpiQ0}. As anticipated in Sec.~\ref{sec:qtm}, the quark PDF becomes
negative at $x \sim 1$. This domain of negative support occurs regardless of the pion body PDF used,
and the reference body PDF actually presents worse behavior in this regard than the NJL model, as the full PDF becomes negative at smaller values of $x$ in this case. 
%If positivity is used to determine the domain of validity for the QTM, then the QTM has a larger domain of validity when paired with the NJL model than when paired with the naive reference body PDF.

The fact that the impulse approximation PDFs are negative at large $x$ is inevitable in the QTM, and can be understood in relation to the DGLAP evolution equations~\cite{Altarelli:1977zs}. The DGLAP kernels can be found from the QTM PDFs by differentiating with respect to the renormalization scale. In fact, one can see the familiar form of the leading order DGLAP kernels in the factors multiplying $I_2(w_{q/g}(x))$ in Eqs.~\eqref{eq:tri_Feyn}, \eqref{eqn:fWqQ}, and \eqref{eqn:fgQ}. The DGLAP kernels, when convoluted with a PDF, result in a function which is negative for $x$ near 1\emdash which is necessary for the evolved PDF to decrease with evolution at $x\sim 1$. It is therefore inevitable that the QTM combined with
the impulse approximation will produce a negative PDF for hadrons. One must go beyond the impulse approximation for the hadron PDFs to be positive-definite.

%-----------------------------------------------------------
%-----------------------------------------------------------

%-----------------------------------------------------------
\begin{figure}[tbp]
  \centering
  \includegraphics[width=\columnwidth]{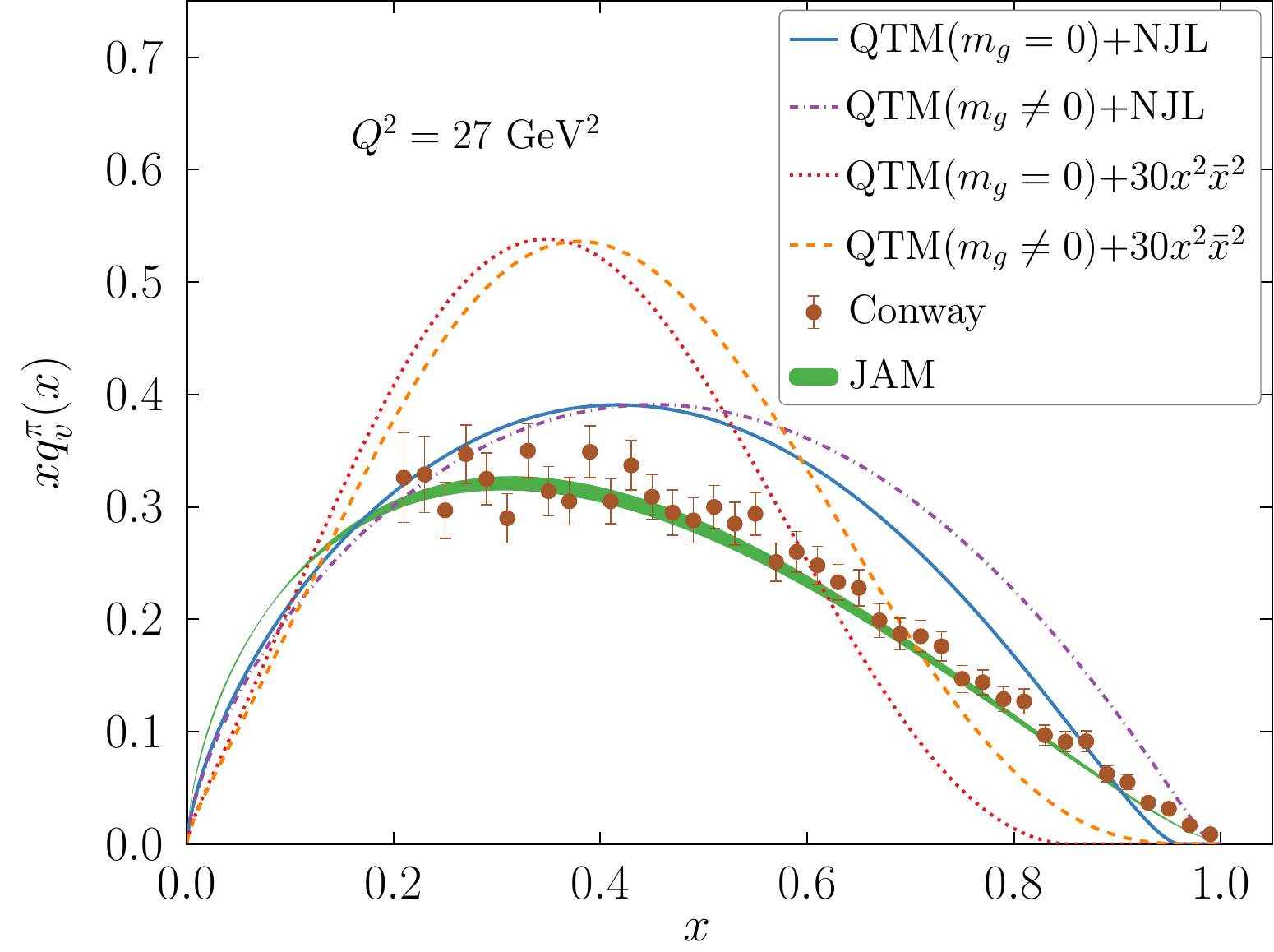}  \\[1.0em]
  \includegraphics[width=\columnwidth]{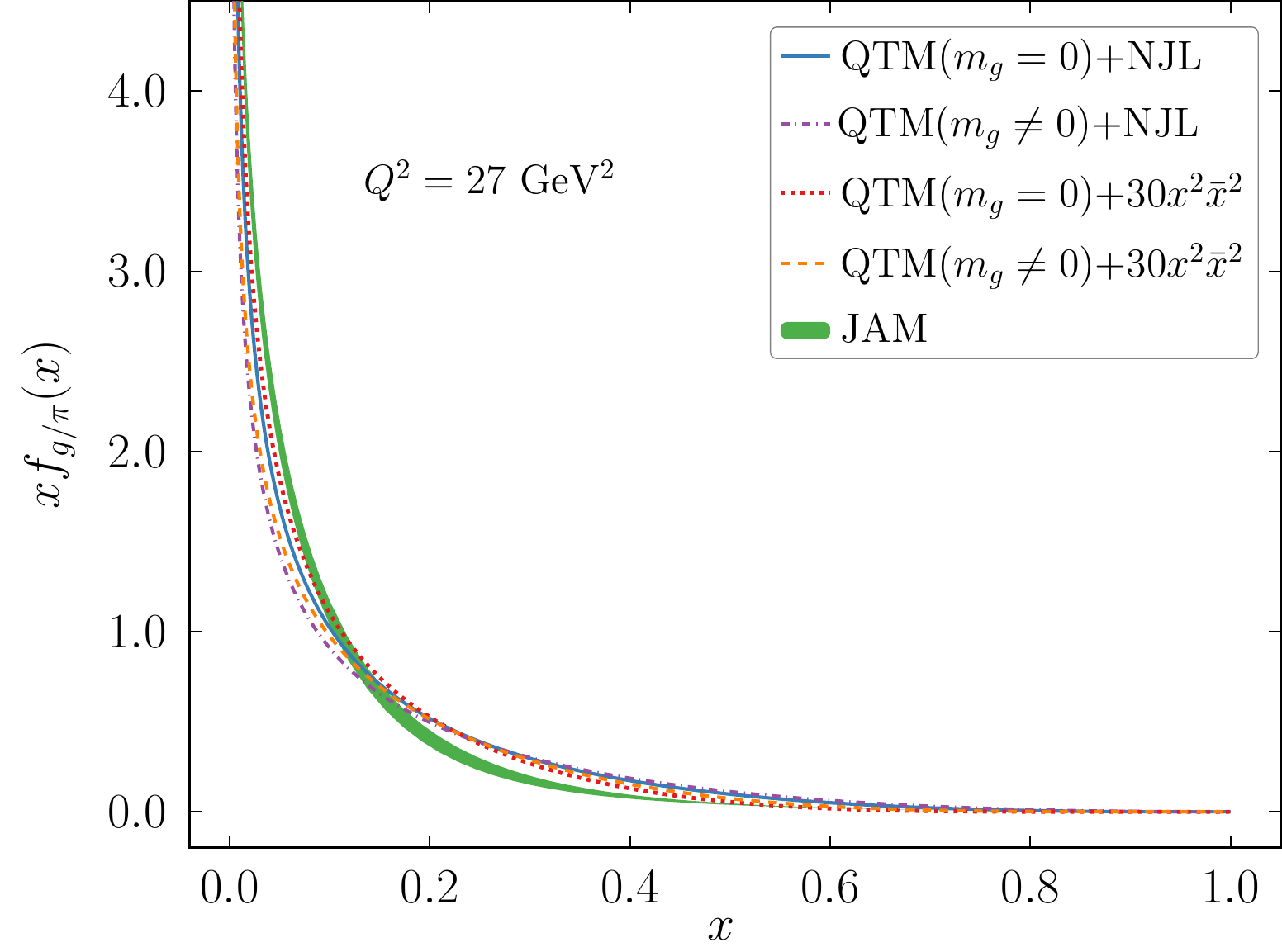}
  \caption{The results presented in Fig.~\ref{fig:pdf:xfpiQ0} evolved from the model scale to $Q^2 = 27\,$GeV$^2$. These results are contrast with the JAM collaboration results from Ref.~\cite{Barry:2018ort} and the Conway {\it et al.} data~\cite{Conway:1989fs}.
%
%Quark (top panel) and gluon (bottom panel) PDFs of the pion at the scale of $Q^2=27\,$GeV$^2$, where we have used the NJL model result (with $m_g=0$ and with a gluon mass $m_g=0.4\,$GeV) and $30\,x^2(1-x)^2$ for pion's body PDFs.
  }
  \label{fig:pdf:qPion}
\end{figure}
%-----------------------------------------------------------

In order to effectively use the DGLAP equations to evolve these results, we set the negative values of the pion's quark PDF to zero and re-weight the quark and gluon distributions to maintain the baryon number and momentum sum rules. We evolve these model PDFs to the empirically relevant scale of $Q^2=27\,$GeV$^2$ using NLO DGLAP equations, and compare to experimental Drell-Yan data from the E615 experiment~\cite{Conway:1989fs}, and the phenomenological parametrization from the JAM analysis~\cite{Barry:2018ort}. These results are given in Fig.~\ref{fig:pdf:qPion}. The first thing to be noticed is that while interesting as a concept to be incorporated at the lagrangian level based on phenomenological ground, we found that a finite gluon mass makes discrepancy with data worse. 

For zero gluon mass, we find that our gluon PDF result has good agreement with the JAM result but the result for the pion valence quark PDF is less satisfactory. The discrepancy with data is an indication of the need to go beyond some of the approximations used in the QTM. Improvements include keeping the quarks off mass shell and thereby removing the impulse approximation. In addition, in the pion there can be gluon exchange between the dressed quark and anti-quark at order $\alpha_s$ and the gluon from the Wilson line can also couple to the spectator quark. These improvements should remove the domain of negative support in the quark PDF. Making these improvements represents a significant calculation and is therefore left for future work.

%%%%%%%%%%%%%%%%%%%%%%%%%%%%%%%%%%%%%%%%%%%%%%%%%%%%%%%%%%%%%%%%%%%%%%%%%%%%%%%%
\section{SUMMARY AND OUTLOOK\label{sec:conc}}
We have constructed a QTM at leading order in the quark-gluon coupling strength. Quark and gluon PDFs of the quark target were calculated directly, including Wilson line contributions in covariant gauges. Gauge invariance of the results was demonstrated by explicitly calculating the PDFs in both covariant and light cone gauges. The quark target PDFs were then combined with a NJL model result and a phenomenological parameterization for the pion PDF via a convolution formalism, and the resulting quark and gluon PDFs were evolved via NLO DGLAP equations and compared to empirical and phenomenological results. We found good agreement between the empirical and calculated gluon PDFs, however, the agreement for the quark PDF is less satisfactory.

This study has produced several interesting results. Perhaps most significant is the finding that the Wilson line can make sizeable contributions to the quark PDFs in covariant gauges, providing as much as a 20\% correction to the quark momentum fraction. Therefore, approaches that do not include the Wilson lines may be failing to account for a large contribution to the quark lightcone momentum. Following indications from lattice QCD that gluons may acquire an effective mass at low-momentum, we explored two methods of including effects from an explicit gluon mass in these calculations. A naive mass term for the gluons is known to violate gauge invariance and we also find that such a term violates the momentum sum rule. We also studied the gluon mass term proposed by Cornwall~\cite{Cornwall:1981zr}, which maintains gauge invariance by also introducing an auxiliary (theta) field. This mass term for the gluon field is found to observe the momentum sum rule as the auxiliary theta-field carries gluon momentum. While exploring the impact of a explicit gluon mass is theoretically interesting -- where we found that a gluon mass reduces the gluon lightcone momentum fraction -- including such a mass term did not provide improved agreement with existing data.

When using the QTM as a method to include intrinsic gluons into low-energy effective theories one shortcoming was identified. This is the unavoidable domain of negative support in the quark PDFs at $x\sim 1$. In covariant gauges this is directly connected to the bilocal operator that defines the PDF and the associated Wilson line contribution. Likely, this can be attributed to the limited way in which gluons were incorporated in the model. For observables defined by local operators this shortcoming will not materialize. The general challenge remains however: How can gluon degrees of freedom be best incorporated in low-energy effective theories of QCD? In principle, the inclusion of gluons into an effective model with just quarks\emdash even if the gluons are only incorporated at leading order\emdash will have effects not only on the inner structure of the dressed quarks, but will also modify the hadron wave functions (e.g. their Bethe-Salpeter amplitudes). After all, introducing a gluon into the model will now make gluon exchange part of the force that binds quarks into hadrons. The QTM+convolution approach taken in this work does not account for gluon exchange between the quarks. Including these effects may provide the necessary ingredients to include intrinsic gluons in effective quark theories and avoid the single shortcoming found in this study.

%%%%%%%%%%%%%%%%%%%%%%%%%%%%%%%%%%%%%%%%%%%%%%%%%%%%%%%%%%%%%%%%%%%%%%%%%%%%%%%%
\begin{acknowledgments}
  C.S.R.C. was supported by Coordena{\c c}{\~a}o de Aperfei{\c c}oamento de Pessoal
  de N{\'i}vel Superior - CAPES, Grants no. 88887.363682 and 88882.330776.
  AF and IC were supported by the U.S.~Department of Energy, Office of Science,
  Office of Nuclear Physics, contract no.~DE-AC02-06CH11357,
  and an LDRD initiative at Argonne National Laboratory
  under Project~No.~2020-0020.
  AF was also supported by U.S.~Department of Energy, Office of Science,
  Office of Nuclear Physics grant no.~DE-FG02-97ER-41014.
  B.E. was supported by Funda\c{c}\~ao de Amparo \`a Pesquisa do Estado de S\~ao Paulo, grant no.~2018/20218-4, and Conselho Nacional de Desenvolvimento Cient\'ifico e Tecnol\'ogico, grant no.~428003/2018-4. 
  G.K was supported in part by: Conselho Nacional de Desenvolvimento Cient\'{\i}fico e Tecnol\'ogico - CNPq, grants no.~309262/2019-4, 464898/2014-5 (INCT F\'{\i}sica  Nuclear e Apli\-ca\-\c{c}\~oes), Funda\c{c}\~ao de Amparo \`a Pesquisa do Estado de S\~ao Paulo - FAPESP, grant no.~2018/25225-9. P.C.T. was supported by the National Science Foundation, grant no.~NSF-PHY1516138.
  CSRC would like to acknowledge the great hospitality of the Theory Group in the Physics Division at Argonne National Laboratory where most of the calculations in this work were carried out.
\end{acknowledgments}

%%%%%%%%%%%%%%%%%%%%%%%%%%%%%%%%%%%%%%%%%%%%%%%%%%%%%%%%%%%%%%%%%%%%%%%%%%%%%%%%
\appendix
\vspace{1.0em}
\section{REGULARIZED INTEGRALS}
\label{sec:integrals}

We use the following notation for basic regularized integrals:
\begin{align}
  I_n(w)
  & \equiv
  i
  \int \frac{\mathrm{d}^4k}{(2\pi)^4}\
  \frac{1}{[k^2-w]^n},
  \label{eqn:In}
\end{align}
to keep the results general and avoid committing to a particular regularization scheme. The specific values of $n=2$ and $n=3$ appear, where the former is logarithmically divergent and the latter convergent. Specific results in several common regularization schemes follow.

In the proper time regularization scheme these basic integrals become:
\begin{align}
  I_2(w)
  &=
  -
  \frac{1}{16\pi^2} \ 
  \Gamma\left(
    0,
    w/\Lambda_{\mathrm{UV}}^2,
    w/\Lambda_{\mathrm{IR}}^2
    \right),
  \\
  I_3(w)
  &=
  \frac{1}{32\pi^2} \frac{1}{w}
  \Big(
    e^{-w/\Lambda_{\mathrm{UV}}^2} - e^{-w/\Lambda_{\mathrm{IR}}^2}
    \Big)
  \,,
\end{align}
where $\Gamma(s,a,b) = \int_a^b \mathrm{d}t\,t^{s-1}\,e^{-t}$ is the generalized incomplete gamma function. Using implicit regularization methods such as constrained differential regularization~\cite{delAguila:1998nd} and constrained implicit regularization~\cite{Battistel:1998sz} gives the results
\begin{align}
  I_2(w)
  &=
  \frac{1}{16\pi^2}
  \log\left(\frac{w}{\Lambda^2}\right), &
  I_3(w)
  &=
  \frac{1}{32\pi^2} \frac{1}{w}.
\end{align}
When using dimensional regularization it is generally important to work in $d=4-2\epsilon$ dimensions from the outset, however, for these simple calculations we can make the replacements
\begin{align}
  I_2(w)
  &=
  -\frac{1}{16\pi^2}
  \left[
  \epsilon^{-1}
  - \log\left(w/\mu^2\right)
  + \log(4\pi) - \gamma_E
  \right],
  \\
  I_3(w)
  &=
  \frac{1}{32\pi^2} \frac{1}{w},
\end{align}
to obtain results in dimensional regularization. The terms $(\epsilon^{-1} + \log(4\pi) - \gamma_E)$ can be removed from $I_2(W)$ by the common $\overline{\text{MS}}$ subtraction scheme.

%%%%%%%%%%%%%%%%%%%%%%%%%%%%%%%%%%%%%%%%%%%%%%%%%%%%%%%%%%%%%%%%%%%%%%%%%%%%%%%%

%\bibliographystyle{apsrev4-2}
\bibliography{main.bib}

\end{document}